\documentclass{svjour3}

\usepackage{amsfonts}
\usepackage{float}
\usepackage{amstext}
\usepackage{amsmath}
\usepackage{amssymb}
\usepackage{color} 
\usepackage{hyperref}
\usepackage{graphicx}
\usepackage[suffix=]{epstopdf}

\newcommand{\NN}{\mathbb{N}}

\newcommand{\RR}{\mathbb{R}}
\newcommand{\ZZ}{\mathbb{Z}}

\newcommand{\EE}{\mathbb{E}}

\title{Consensus Convergence with Stochastic Effects}
\dedication{Dedication to Willi J\"{a}ger's 75th Birthday}
\author{Josselin Garnier \and George Papanicolaou \and Tzu-Wei Yang}
\institute{Josselin Garnier \at Laboratoire de Probabilit\'{e}s et Mod\`{e}les Al\'{e}atoires \& Laboratoire Jacques-Louis Lions, Universit\'{e} Paris Diderot, 75205 Paris Cedex 13, France\\
\email{garnier@math.univ-paris-diderot.fr}
\and 
George Papanicolaou \at Department of Mathematics, Stanford University, Stanford, CA 94305, USA \\ \email{papanicolaou@stanford.edu}
\and
Tzu-Wei Yang \at School of Mathematics, University of Minnesota, Minneapolis, MN 55455, USA \\ \email{yangx953@umn.edu}
}

\date{}

\begin{document}
\maketitle

\begin{abstract}
We consider a stochastic, continuous state and time opinion model where each agent's opinion locally interacts with other agents' opinions in the 
system, and there is also exogenous randomness.  The interaction tends to create clusters of common opinion.
By using linear stability analysis of the associated nonlinear Fokker-Planck equation that governs the empirical density of opinions
in the limit of infinitely many agents, 
we can estimate the number of clusters, the time to 
cluster formation and the critical strength of randomness so as to have cluster formation.  We also discuss the cluster dynamics after their formation, the 
width and the effective diffusivity of the clusters. Finally, the long term behavior of clusters is explored numerically.  Extensive numerical simulations confirm our analytical findings.


\end{abstract}
\section{Introduction}
\label{sec:intro}

Opinion dynamics models have attracted a lot of attention and there are many analytical and numerical  studies that consider different models arising
from many different fields. 
In much of the literature, an opinion dynamics model is a system with a large number of opinion variables, $x_i(t)$, $i=1,\ldots,N$, taking values in  $\mathbb{R}^n$.  The time evolution of the opinion variables is governed by an attractive interaction between any two opinion variables, often taken to be a nonnegative function of the Euclidean distance of the two opinion variables 
and may also be time dependent. The most interesting feature of such a model is that opinions only interact \textit{locally} and the influence 
function is compactly supported, interpreted as bounded confidence.  In this case, it is of interest to know whether the system will exhibit  
\textit{consensus convergence}, which means that all the opinion variables converge to the same point as time tends to infinity. Except for some specific 
consensus models, a broad sufficient condition to have consensus convergence for a general class of models is not known.  However, several studies 
have shown that for a variety of different types of consensus building interactions, and without external forces or randomness, the opinions will converge 
to possibly several clusters. In this case, the distance between 
distinct clusters should be larger than the support of the influence region. But it is not known, in general, how to determine the number of clusters.

A more realistic way to model opinion dynamics is to add external randomness to the system.  In this case, the model becomes a system of $N$ 
stochastic processes and usually the randomness in the model is independent from one opinion holder or agent to another. Many deterministic techniques can also
be used in the stochastic case, but some methods, such as the use of master equations, are particularly 
useful in stochastic models. When the external noise is large in the stochastic models then the tendency to cluster is
effectively eliminated as the system is dominated by the noise. This is a phenomenon seen elsewhere in statistical physics as well.
The strength of the noise or randomness must be below a critical value in order for cluster formation to emerge and evolve.

The literature in opinion dynamics is very extensive so we mention only a few papers that have guided our own work.  Hegselmann and Krause   
\cite{Hegselmann2002} consider a discrete-time evolution model, in which the opinions in the next step are the average of the current opinions within a specified range of 
the influence region. Pineda \textit{et al.} \cite{Pineda2013} add noise to the Hegselmann-Krause model and determine the critical strength of the noise so as to have cluster 
formation, using a master equation approach and linear stability analysis.  The same method is also used in \cite{Pineda2009}\cite{Pineda2011} on the 
Deffuant-Weisbuch model \cite{Deffuant2000}.  In \cite{Canuto2012}, the authors take the limit as the number of opinions goes to infinity and consider the 
distribution of the opinions (the Eulerian approach), instead of tracking every single opinion in the Hegselmann-Krause model (the Lagrangian approach), and 
\cite{Mirtabatabaei2014} further discuss the case with external forces.  The long time behavior and a sufficient condition for consensus convergence of the 
Hegselmann-Krause model are considered in \cite{Blondel2005}\cite{Blondel2007}\cite{Yang2012}. The long time behavior of the Hegselmann-Krause model with a general influence function is discussed in \cite{Jabin2014}\cite{Motsch2014}.  The Hegselmann-Krause model involving different types of agents is considered in \cite{Hegselmann2015}.
Some recent development of the study of opinion dynamics are in \cite{LORENZ2007}\cite{Motsch2014}. Other, related relevant works are
\cite{Ha2008}\cite{Duering2009}\cite{Maes2010}\cite{Como2011}\cite{GOMEZ-SERRANO2012}\cite{Carro2013}\cite{Huang2013}\cite{Lanchier2013}\cite{Baccelli2014}.

Our contributions in this paper are the following.  We consider a stochastic opinion model where every opinion is influenced by an independent Brownian 
motion. By the mean field limit theory, the empirical probability measure of the opinions converges, as the size of the population goes to infinity, to a 
solution of a nonlinear Fokker-Planck equation.  Using a linear stability analysis, we estimate the number of clusters, the time to cluster formation and the 
critical strength of the Brownian motions to have cluster formation.  The linear stability analysis can be applied to both deterministic and stochastic models. We 
also discuss the behavior of the system after the initial cluster formation but before further cluster consolidation, where the centers of 
the clusters are expected to behave like independent Brownian motions.  Finally, we consider the long time behavior of the system. 
Once clusters are formed, their centers behave like Brownian motions until further merging. 
After consensus convergence, where there is only one cluster, there is a small probability that all the opinions inside the limit cluster 
will spread out and the system will become an independent 
agent evolution.  Extensive numerical simulations are carried out to support our analysis and remarks about cluster formation and evolution.

The paper is organized as follows. The interacting agent model is presented in section \ref{sec:model}. The mean field limit is presented
briefly in section \ref{sec:mf}. The linearized stability analysis of the governing nonlinear Fokker-Planck equation is presented in section \ref{sec:det}
when there is no external noise. The results of numerical simulations are also presented in this section. In section \ref{sec:stoch} we extend
the analysis of the previous section to the stochastic case when there are external noise influences. We also present the results of numerical
simulations in the stochastic case. In section \ref{sec:long} we comment briefly about the long time behavior in the stochastic case when there is
clustering. We end with a brief summary and conclusions in section \ref{sec:conclusion}.
\section{The interacting agent model}
\label{sec:model}

The opinion model we consider in this paper is (see  \cite[Eq. (1.2a)]{Motsch2014}):
\begin{equation}
	\label{eq:dx_i}
	dx_i = -\frac{1}{N} \sum_{j=1}^N a_{ij} (x_i-x_j) dt+ \sigma dW^i(t),
\end{equation}
where $x_i(t)$ is the agent $i$'s opinion modeled as real valued process, where $t$ is time and $i=1,\ldots, N$. The coefficients $a_{ij}$ denote the strength of the interaction between 
$x_i$ and $x_j$ and they are a function of the distance between $x_i$ and $x_j$:
\begin{equation}
	\label{eq:a_ij}
	a_{ij} = \phi( |x_i-x_j |).
\end{equation}
The interesting case is when the influence function $\phi$ is non-negative and compactly supported. In other words, the interactions are attractive and the agent $i$ 
affects only the other agents that have similar opinions. Here we assume that $\phi$ is compactly supported in $[0, R_0]$
\begin{equation}
	\label{eq:phi_0}
	\phi(r) =\phi_0\left(\frac{r}{R_0} \right)
\end{equation}
where  $\mathrm{supp}(\phi_0)=[0,1]$.

The $W^i(t)$, $i=1,\ldots,N$ are independent standard Brownian motions that model the uncertainties of the agents' opinions, and $\sigma$ is a 
non-negative constant quantifying the strength of the uncertainties. If $\sigma=0$, then there is no randomness in this model and (\ref{eq:dx_i}) is a 
deterministic system, while if $\sigma>0$, the system becomes stochastic.

For the purposes of the analysis below, we consider the model (\ref{eq:dx_i}) on the torus $[0,L]$ instead of the real line $\RR$. i.e. we consider 
the model in the bounded space $[0,L]$ with periodic boundary conditions. The assumption of periodic boundary conditions is mostly for simplifying 
the analysis. Although this assumption may not be appropriate in some applications, we found that the results obtained using it
are numerically consistent with the same model in full space or in a finite interval with reflecting boundary conditions. The later two 
are in general more realistic for many applications. We note that the same periodicity assumption for the analysis of the opinion dynamics is also used 
in \cite{Pineda2009}\cite{Pineda2011}\cite{Pineda2013}.
\section{The mean field limit}
\label{sec:mf}

At time $t$, we consider the empirical probability measure $\rho^N(t,dx)$ of the opinions of all the agents:
\begin{equation}
	\rho^N(t,dx) = \frac{1}{N} \sum_{j=1}^N \delta_{x_j(t)}(dx).
\end{equation}
Here $\delta_x(dx)$ is the Dirac measure with the point mass at $x$. The empirical probability measure $\rho^N(t,dx)$ is a
measure valued stochastic process. 
We assume that as $N\to\infty$, $\rho^N(0,dx)$ converges weakly,
in probability to $\rho_0(dx)$ which is a deterministic measure with density $\rho_0(x)$.
By using the well known mean field asymptotic theory (see, for example, 
\cite{Dawson1983}\cite{Gartner1988}\cite{Kurtz1999}), it can be shown that as $N\to\infty$, $\rho^N(t,dx)$ converges weakly,
in probability to $\rho(t,dx)$, for $0\leq t\leq T<\infty$, a deterministic probability measure. Under suitable conditions the limit measure 
has a density $\rho(t,x)$ which satisfies (in a weak sense) the nonlinear
Fokker-Plank equation:
\begin{equation}
	\label{eq:Fokker-Plank equation}
	\frac{\partial \rho}{\partial t} (t,x)
	= \frac{\partial}{\partial x} \left\{ \left[ \int \rho(t,x-y) y \phi(|y|) dy \right]\rho(t,x) \right\} 
	+ \frac{\sigma^2}{2} \frac{\partial^2 \rho}{\partial x^2}(t,x),
\end{equation}
with given initial density $\rho_0(x)$.
In particular, if $x_1(0),\ldots,x_N(0)$ are sampled independently and identically according to the uniform measure over $[0,L]$, then the result 
holds true and the initial measure has constant density $\rho_0(x) = 1/L$.

In this paper, we assume that $N$ is large and view the mean field limit as the defining problem. Therefore, we will analyze the overall behavior 
of the opinion dynamics, $x_1(t),\ldots,x_N(t)$, by analyzing instead the nonlinear Fokker-Planck equation (\ref{eq:Fokker-Plank equation}).
\section{Deterministic consensus convergence: $\sigma=0$}
\label{sec:det}

We will follow a modulational instability approach to study the mean field limit when ($\sigma=0$), also analyzed in 
\cite{Motsch2014}\cite{Jabin2014}. We look for conditions so as to have consensus convergence where all the opinions converge to a cluster as $t\to\infty$.
We also analyze the number of clusters if there is no consensus convergence and the time to cluster formation, that is, the onset of cluster formation.

\subsection{Linear stability analysis}

We first linearize the Fokker-Planck equation (\ref{eq:Fokker-Plank equation}) with $\sigma=0$ by assuming that 
$\rho(t,x) = \rho_0 + \rho_1(t,x) = 1/L + \rho_1(t,x)$. Substituting $\rho(t,x) = \rho_0 + \rho_1(t,x)$ into (\ref{eq:Fokker-Plank equation}) and 
assuming that $\rho_1$ is a small perturbation of $\rho$ so that the $O(\rho_1^2)$ term is negligible, we find that $\rho_1(t,x)$ satisfies:
\begin{align}
	\label{eq:linearized mean-field with zero sigma}
	\frac{\partial \rho_1}{\partial t}(t,x) 
	&= \frac{\partial}{\partial x}\left[\int \rho_0y\phi(|y|)dy \rho_1(t,x)\right]
	+ \frac{\partial}{\partial x}\left[\int \rho_1(t,x-y)y\phi(|y|)dy \rho_0\right]\\
	&= \rho_0 \int \frac{\partial \rho_1}{\partial x}(t,x-y) y \phi(|y|) dy. \notag
\end{align}
The last equality in (\ref{eq:linearized mean-field with zero sigma}) holds because $\phi(|y|)$ is an even function and therefore 
$\int y\phi(|y|)dy=0$.

By taking the Fourier transform in $x$, $\hat{\rho}_1(t,k)=\int_0^L \rho_1(t,x) e^{-ikx} dx$, with the discrete set of frequencies $k$ in 
\begin{equation}
	\mathcal{K} = \{2\pi n /L, n \in \ZZ\},
\end{equation}
we find from (\ref{eq:linearized mean-field with zero sigma}) that
\begin{equation}
	\label{eq:linearized mean-field with zero sigma, Fourier transform}
	\frac{\partial \hat{\rho}_1}{\partial t} (t,k) = \left[ i \rho_0 k \int e^{-iky} y \phi(|y|) dy \right] \hat{\rho}_1(t,k),
\end{equation}
which gives the growth rates of the modes:
\begin{equation}
	\label{eq:gamma_k}
	\gamma_k = \mathrm{Re} \left[ i \rho_0 k \int e^{-iky} y \phi(|y|) dy \right] = \rho_0 k \int \sin(ky) y \phi(|y|) dy.  
\end{equation}
We can see that for each $k$, $|\hat{\rho}_1(t,k)| = |\hat{\rho}_1(0,k)|\exp(\gamma_k t)$. 
By replacing $\phi$ with $\phi_0$ (see (\ref{eq:phi_0})), we can rewrite $\gamma_k$ as
\begin{equation}
	\label{eq:psi}
	\gamma_k = \rho_0 R_0 \psi( kR_0 ),\quad \psi(q) = 2q \int_0^1 \phi_0(s) s \sin(q s) ds.
\end{equation}
The growth rate $\gamma_k$ is maximal for $k = \pm k_{\max} $ with $k_{\max} := q_{\max}/R_0$, more exactly, for $k$ equal to 
plus or minus the discrete frequency $k_{\max}$ in the set 
${\cal K}$ that maximizes $\psi(kR_0)$, which is close to $q_{\max}/R_0$. Here 
\begin{equation}
	\label{eq:qmax}
	q_{\max} = \underset{q>0}{\arg\max} \left[\psi(q)\right] = \underset{q>0}{\arg\max} \left[2q \int_0^1 \phi_0(s) s \sin(q s) ds\right]
\end{equation}
The optimal (continuous)  frequency $q_{\max}$ is positive and finite under general conditions since 
$\psi(q) \simeq 2 q^2 \int_0^1 s^2 \phi_0(s) ds$ for $0\leq q \ll 1$ and $\psi(q)$ is bounded or decays to zero at infinity depending on the 
regularity of $\phi_0$.

\subsection{Fluctuation theory}
By the central limit theorem, if we assume that the initial opinions $x_1(0), \ldots, x_N(0)$ are sampled identically according to the uniform 
distribution over the domain $[0,L]$, then 
\begin{equation*}
	\rho^N_1(t=0,dx) :=  \sqrt{N} \left( \rho^N(t=0,dx) - \rho_0(dx) \right)
	= \sqrt{N} \left(\frac{1}{N} \sum_{i=1}^{N}\delta_{x_i(0)}(dx) - \frac{dx}{L} \right)
\end{equation*}
converges in distribution as $N \to \infty$ to the measure $\rho_1(t=0,dx)$, whose frequency components, for $k \in {\cal K}\backslash \{0\}$,
\begin{align*}
	\hat{\rho}_1(t=0,k) 
	&= \lim_{N\to\infty}\int_0^L \sqrt{N}  e^{-ikx} \left(\frac{1}{N}\sum_{j=1}^{N}\delta_{x_j(0)}(dx)-\frac{dx}{L}\right)\\
	&= \lim_{N\to\infty} \sqrt{N} \left( \frac{1}{N}\sum_{j=1}^{N} e^{-ikx_j(0)} - \int_0^L \frac{1}{L}e^{-ikx}dx \right)\\
	&= \lim_{N\to\infty} \sqrt{N} \left( \frac{1}{N}\sum_{j=1}^{N} e^{-ikx_j(0)}\right)
\end{align*}
are independent and identically distributed with complex circular Gaussian random variables with mean zero and variance $1$:
\begin{equation*}
	\EE \left[ \hat{\rho}_1(t=0,k) \right]=0,\quad
	\EE \left[ \hat{\rho}_1(t=0,k) \overline{\hat{\rho}_1(t=0,k')} \right] = \delta_{kk'},\quad 
	k,k' \in {\cal K} \backslash \{0\},
\end{equation*}
 $\hat{\rho}_1(t=0,-k) = \overline{ \hat{\rho}_1(t=0,k) }$, while $\hat{\rho}_1(t=0,k=0) =0$.
 
For any $T$, the measure-valued process 
\begin{equation}
	\rho^N_1(t,dx) :=  \sqrt{N} \left( \rho^N(t,dx) - \rho_0(dx) \right) , \quad t \in [0,T]
\end{equation}
converges in distribution as $N \to \infty$ to a measure-valued process $\rho_1(t,dx)$ whose density $\rho_1(t,x)$ satisfies the deterministic PDE
\begin{equation}
	\label{eq:rho1}
	\frac{\partial \rho_1}{\partial t} (t,x) 
	= \rho_0 \int \frac{\partial \rho_1}{\partial x}(t,x-y) y \phi(|y|) dy  
\end{equation}
with the random initial condition described above.

Consequently, combining with (\ref{eq:linearized mean-field with zero sigma, Fourier transform}) and (\ref{eq:gamma_k}), at any time $t$, the 
frequency components $\hat{\rho}_1(t,k)$, $k \in {\cal K} \backslash \{0\}$, are independent complex circular Gaussian random variables, with mean zero and variance 
$\exp (2 \gamma_k t)$:
\begin{equation}
	\EE \left[ \hat{\rho}_1(t,k) \overline{\hat{\rho}_1(t,k')} \right]
	= \delta_{kk'} \exp (2 \gamma_k t), \quad k,k' \in {\cal K} \backslash \{0\},
\end{equation}
$\hat{\rho}_1(t,-k) = \overline{\hat{\rho}_1(t,k)}$, while $\hat{\rho}_1(t,k=0)=0$. Therefore, 
\begin{align*}
	&\EE\left[\rho_1(t,x)\rho_1(t,x')\right]
	= \EE\left[\sum_k\hat{\rho}_1(t,k)\frac{e^{ikx}}{L} \sum_k\hat{\rho}_1(t,k)\frac{e^{ikx'}}{L}\right]\\
	&\quad = \sum_k \EE\left[\hat{\rho}_1(t,k)\frac{e^{ikx}}{L} \hat{\rho}_1(t,-k)\frac{e^{-ikx'}}{L}\right]
	= \frac{1}{L^2} \sum_{k\neq 0} e^{2\gamma_k t} e^{ik(x-x')}.
\end{align*}

For large times, the spectrum of $\rho_1(t,x)$ becomes concentrated around the optimal wavenumber $k_{\max}$. We can expand 
$\gamma_k = \gamma_{\max} + \frac{1}{2} \gamma_{\max}'' (k-k_{\max})^2$ for $k$ around $k_{\max}$ and use a continuum approximation for 
the discrete sum:
\begin{align}
\notag
	&\EE\left[\rho_1(t,x)\rho_1(t,x')\right] 
	= \frac{1}{L^2} \sum_{k\neq 0} e^{2\gamma_k t} e^{ik(x-x')}
	\simeq \frac{1}{L^2} \int_{-\infty}^{\infty} e^{2\gamma_k t} e^{ik(x-x')}\frac{L}{2\pi} dk\\
\notag
	& =\frac{1}{2\pi L} \int_{-\infty}^{\infty} e^{2\gamma_k t} e^{ik(x-x')} dk
	\simeq \frac{1}{\pi L} {\rm Re}\left( \int_{0}^{\infty}
	e^{2(\gamma_{\max}+\frac{1}{2}\gamma_{\max}''(k-k_{\max})^2)t} e^{ik(x-x')}dk \right)\\
\notag
	& =\frac{1}{\pi L}{\rm Re}\left( e^{2\gamma_{\max}t}\int_{0}^{\infty} e^{\gamma_{\max}''(k-k_{\max})^2t}e^{ik(x-x')}dk \right) \\
	&  = \left(\frac{1}{L}e^{2\gamma_{\max}t} \cos ( k_{\max}(x-x')) \right)
	\left(\frac{1}{\sqrt{\pi|\gamma_{\max}''| t}} e^{-\frac{(x-x')^2}{4|\gamma_{\max}''| t}}\right).
	\label{eq:large t covariance 1}
\end{align}

A typical realization of $\rho_1(t,x)$ is a modulation with the carrier spatial frequency $k_{\max}$
and a slowly varying envelope with stationary Gaussian statistics, mean zero, and Gaussian covariance function.
This is valid provided $L^2 \gg 4|\gamma_{\max}''| t$. If $L^2 \ll 4|\gamma_{\max}''| t$, then the 
continuum approximation is not valid and we have
\begin{align}
\notag
	\EE \left[ \rho_1(t,x) \rho_1(t,x') \right] 
	&= \frac{1}{L^2} \sum_{k\neq 0} e^{2\gamma_k t} e^{ik(x-x')}
	= \frac{2}{L^2} \sum_{k>0} e^{2\gamma_k t} \cos(k(x-x'))\\
	&\simeq \frac{2}{L^2} e^{ 2 \gamma_{\max} t}  \cos \left(k_{\max} (x-x')\right).
	\label{eq:large t covariance}
\end{align}
A typical realization of $\rho_1(t,x)$ is a modulation with the carrier spatial frequency $k_{\max}$.

Because $\gamma_{\max}>0$, the linear system (\ref{eq:rho1}) is unstable and therefore the central limit theorem cannot be extended to 
arbitrarily large times. In fact the theorem is limited to times $t$ such that $\rho_1(t,x)/\sqrt{N}$ is smaller than $\rho_0=1/L$ so that the 
linearization around $\rho_0$ is valid. Therefore the time up to the onset of clustering is when the perturbation $\rho_1$ becomes of the same 
order as $\sqrt{N} \rho_0$, that is to say when  $\EE \big[ \rho_1(t_{\rm clu},x)^2 \big] \simeq N L^{-2}$, which is approximately (up to terms 
smaller than $\ln N$):
\begin{equation*}
	t_{clu} \simeq \frac{1}{2 \gamma_{\max} } \ln N  
	\simeq \frac{1}{2 \rho_0 R_0  \psi(q_{\max} )} \ln N 
\end{equation*}
when $N \gg 1$.

We note that the fact that a random initial distribution gives rise to a quasi-deterministic subsequent evolution by spectral gain selection
occurs in many fields, for instance in fluid mechanics (hydrodynamic instabilities) or in nonlinear optics (beam filamentation).

\subsection{Consensus convergence}
The linear stability analysis shows that the opinion dynamics, starting from a uniform distribution of agents, gives clustering with a mean 
distance between clusters equal to $2\pi /k_{\max}$ (see (\ref{eq:large t covariance 1}) and (\ref{eq:large t covariance})). Once clustering has occurred, two types of
dynamical evolutions are possible:
\begin{enumerate}
	\item If $2\pi /k_{\max} > R_0$, then the clusters do not interact with each other because they are beyond the range of the influence 
	function. Therefore, the situation is frozen and there is no consensus convergence.
	
	\item If $2\pi /k_{\max} < R_0$, then the clusters interact with each other. There may be consensus convergence. However, consensus 
	convergence is not guaranteed as clusters may merge by packets, and the centers of the new clusters may be separated by a distance larger than 
	$2\pi / k_{\max}$, and then global consensus convergence does not happen. The number of mega-clusters formed by this dynamic is not easy to 
	predict.
\end{enumerate}
If we neglect the rounding and consider $k_{\max} = q_{\max} /R_0$, which is possible if $q_{\max} L/R_0 \gg 1$, then the criterion 
$2\pi /k_{\max} >$ or $< R_0$ does not depend on $R_0$, as it reads $2\pi / q_{\max} >$ or $<1$, which depends only on the normalized 
influence function $\phi_0$ by (\ref{eq:psi}) and (\ref{eq:qmax}).

These two dynamics can be observed in the examples of Figure 1.1 in \cite{Motsch2014}:
\begin{enumerate}
	\item If $\phi(r)={\bf 1}_{[0,1]}(r)$, then $q_{\max}\simeq  2.75$ and the mean distance between clusters is about $2.3$, that is beyond the 
	range $1$ of the influence function, and there is no consensus convergence.
	 
	\item If $\phi(r)=0.1\times{\bf 1}_{[0,1/\sqrt{2}]}(r)+{\bf 1}_{[1/\sqrt{2},1]}(r)$, then $q_{\max}\simeq 9.1$ and the mean distance between 
	clusters is about $0.7$, that is within the range of the influence function, and there is consensus convergence.
\end{enumerate}
These predictions are quantitatively in very good agreement with the numerical simulations (distance between clusters and so on). 

To summarize, the main result in the noiseless case $\sigma=0$ is as follows. In the regime $N \to \infty$, there is no consensus convergence if 
$q_{\max} < 2\pi$. There may be consensus convergence if $q_{\max} > 2\pi$. Of course this stability analysis and the result that follow 
can be extended easily to the multi-dimensional case, and to other types of opinions or flocking dynamics.

\subsection{Numerical simulations}

We use the explicit Euler scheme to simulate the deterministic opinion dynamic (\ref{eq:dx_i}) when $\sigma=0$:
\begin{equation}
	\label{eq:discrete dx_i}
	x_i^{n+1} = x_i^n -\frac{1}{N} \sum_{j=1}^N \phi(|x_i^n-x_j^n|) (x_i^n-x_j^n) \Delta t,\quad \phi(s) = \phi_0(s/R_0).
\end{equation}
Although our analysis is on the torus $[0,L]$, we still simulate (\ref{eq:discrete dx_i}) on the full space. The simulation 
results indicate, however, that the analysis under periodic conditions is still consistent with the numerics with different boundary conditions. As it is 
shown in \cite{Motsch2014}\cite{Jabin2014}, if $x_1(0),\ldots,x_N(0)$ are in the interval $[0,L]$, then $x_1(t),\ldots,x_N(t)\in [0,L]$ for any 
$t\geq 0$.

\begin{figure}
	\centering
	\includegraphics[width=0.49\linewidth]{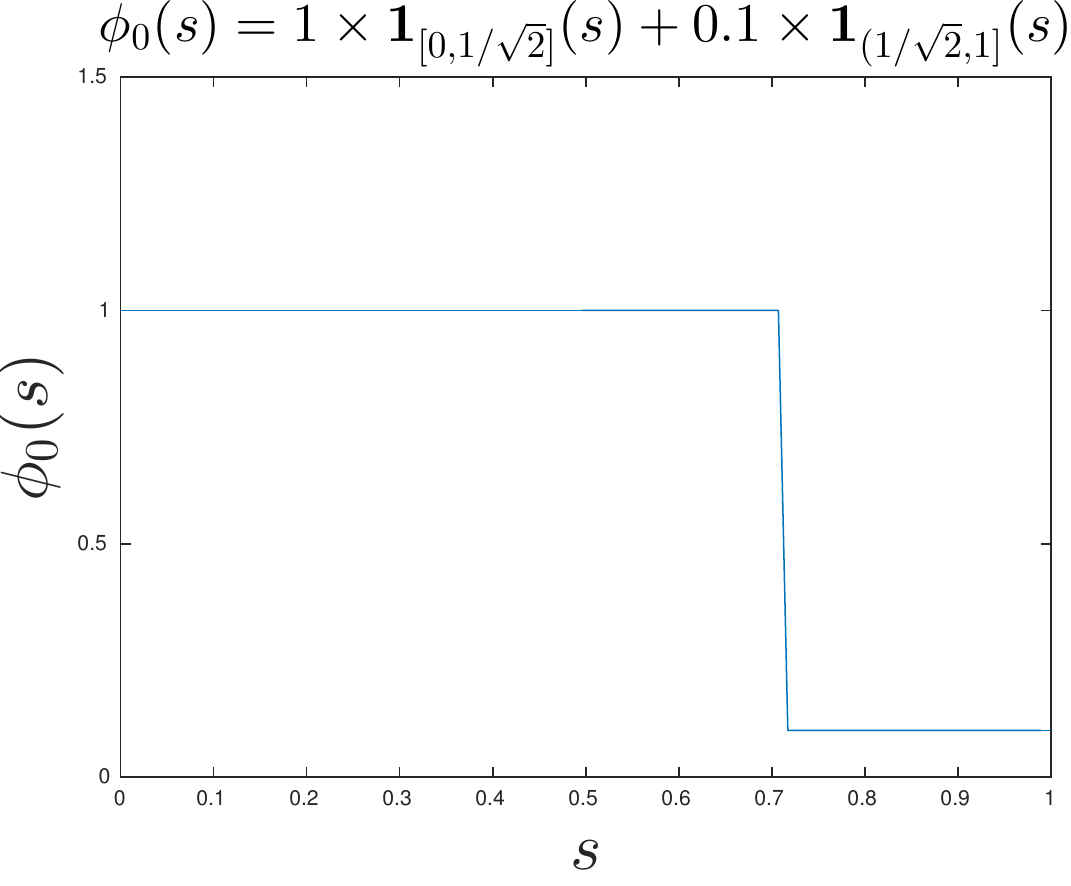}
	\includegraphics[width=0.49\linewidth]{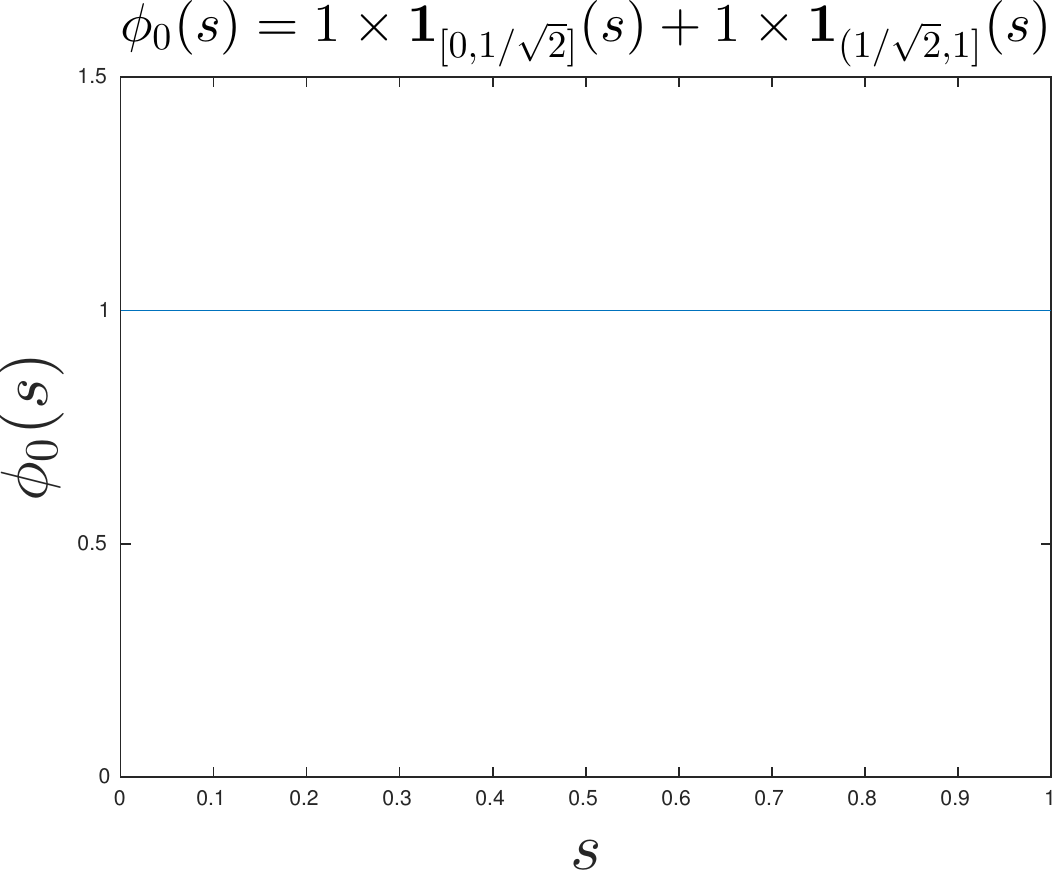}
	
	\includegraphics[width=0.49\linewidth]{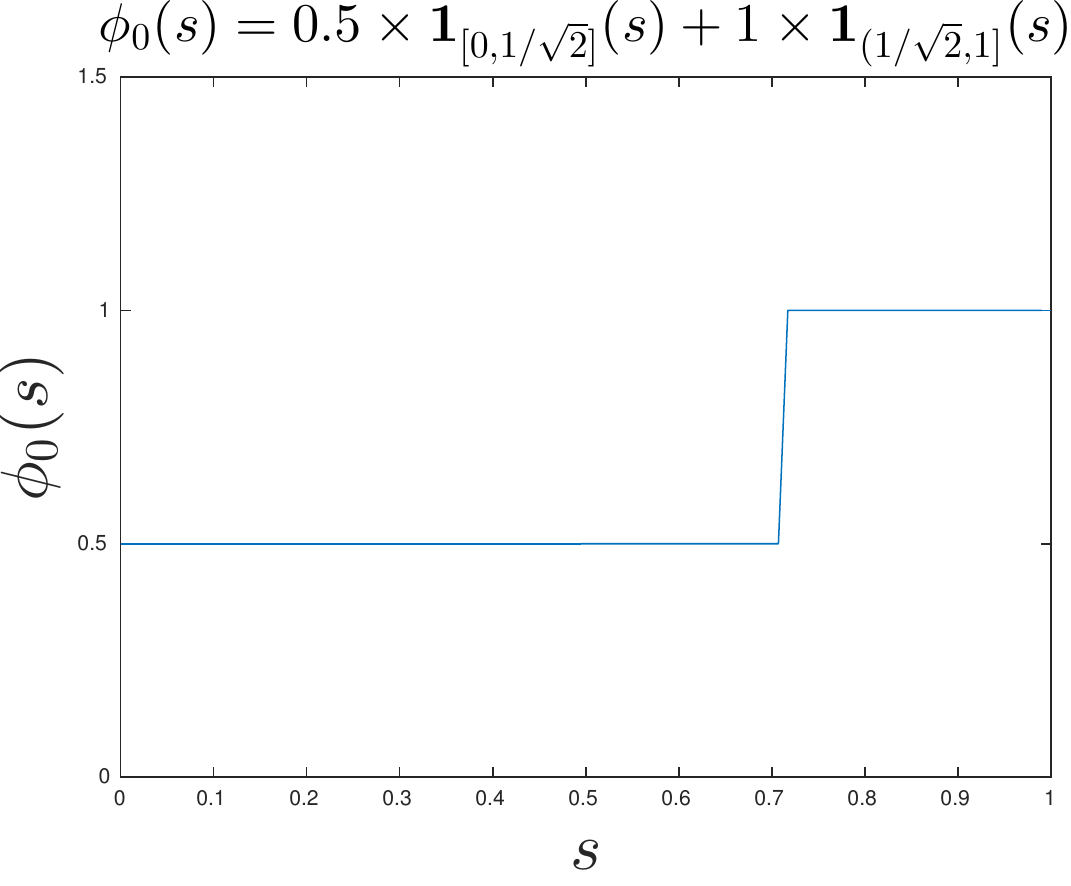}
	\includegraphics[width=0.49\linewidth]{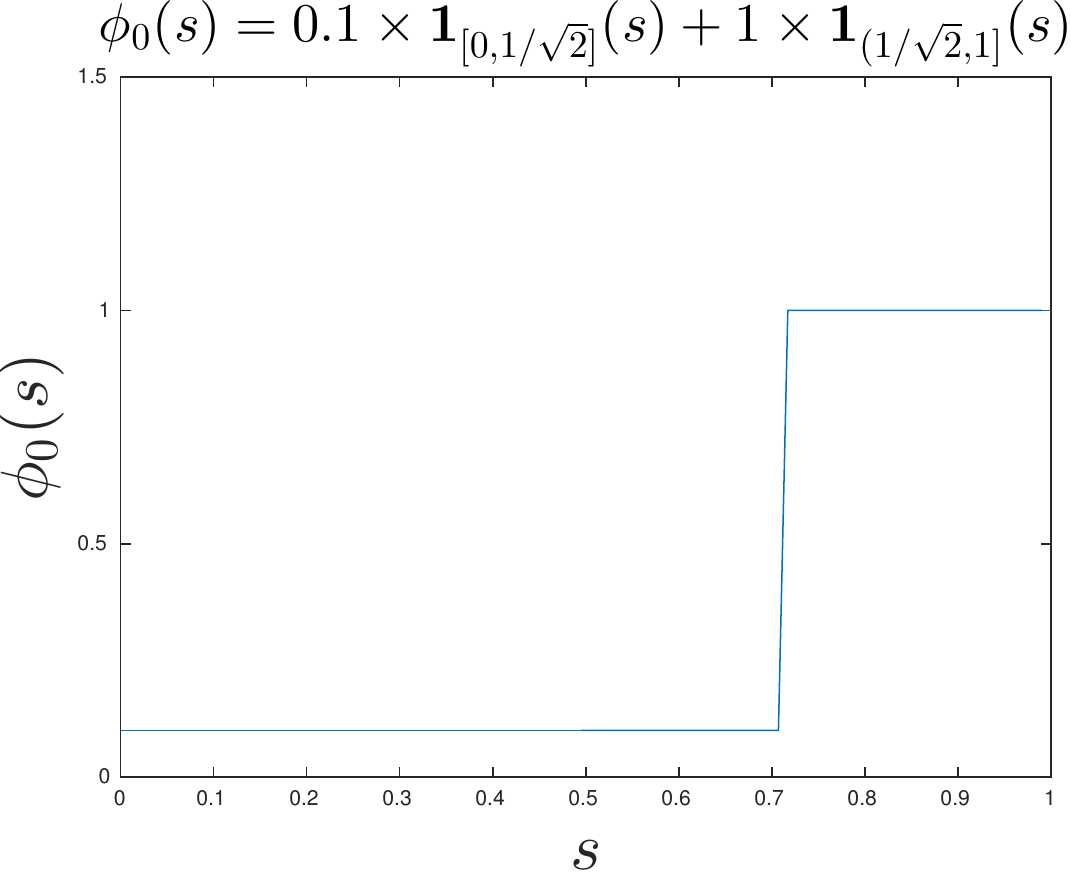}
	
	\includegraphics[width=0.49\linewidth]{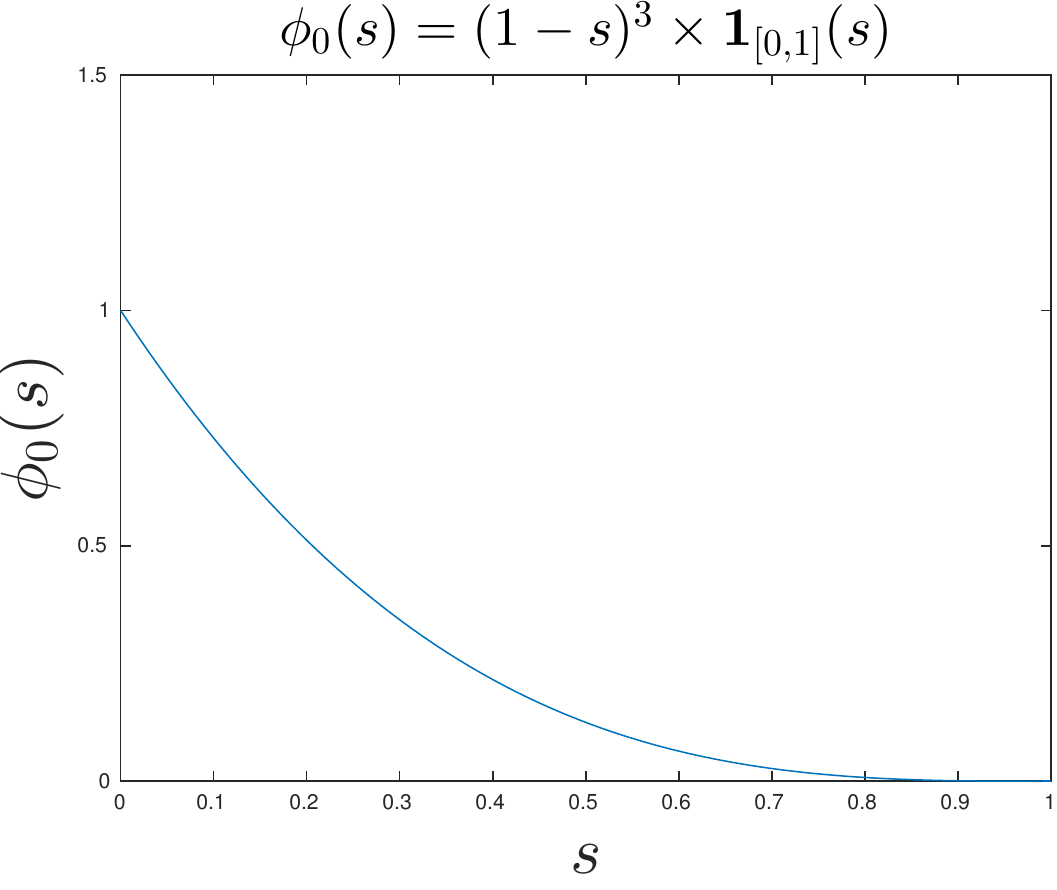}
	\includegraphics[width=0.49\linewidth]{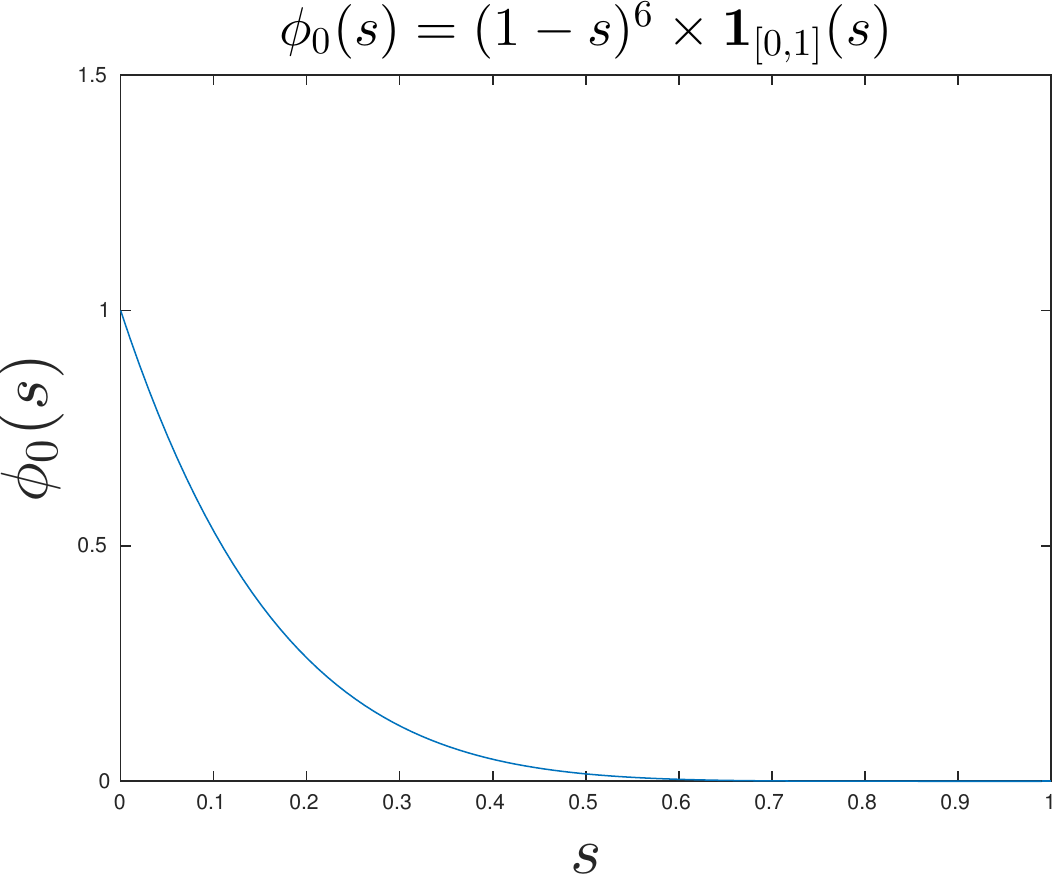}
	
	\caption{The plots of the influence functions $\phi_0(s)$.}
	\label{fig:influence functions}
\end{figure}

We test for the influence functions studied in \cite{Motsch2014}\cite{Jabin2014}:
\begin{align*}
	\phi_0^1(s) &= \mathbf{1}_{[0,1/\sqrt{2}]}(s) + 0.1\times\mathbf{1}_{(1/\sqrt{2}, 1]}(s)\\
	\phi_0^2(s) &= \mathbf{1}_{[0,1]}(s)\\
	\phi_0^3(s) &= 0.5\times\mathbf{1}_{[0,1/\sqrt{2}]}(s) + \mathbf{1}_{(1/\sqrt{2}, 1]}(s)\\
	\phi_0^4(s) &= 0.1\times\mathbf{1}_{[0,1/\sqrt{2}]}(s) + \mathbf{1}_{(1/\sqrt{2}, 1]}(s)\\
	\phi_0^5(s) &= (1-s)^3\times\mathbf{1}_{[0,1]}(s)\\
	\phi_0^6(s) &= (1-s)^6\times\mathbf{1}_{[0,1]}(s),
\end{align*}
and their plots are shown in Figure \ref{fig:influence functions}.

We compute the key quantity $q_\mathrm{max}$ by exploring all possible $q$ in $[0,100]$:
\begin{equation*}
	q_{\max} = \underset{R_0q\in\mathcal{K}, 0<q\leq 100}{\arg\max} \left[2q \int_0^1 \phi_0(s) s \sin(q s) ds\right].
\end{equation*}
We find that for the cases of $\phi_0^3$ and $\phi_0^4$, $q_{\max}$ are not unique and the non-uniqueness of $q_{\max}$ will greatly affect 
the results of the consensus convergence. 
The parameters we use for the simulation are $\Delta t=0.1$, $L=10$, $R_0=1$ and $N=500$. For each $\phi_0$, we also plot the function 
$\psi(s)=2q\int_0^1 \phi_0(s) s \sin(qs) ds$; the stars in the plots are the values of $\psi(s)$ at $R_0q\in\mathcal{K}$ and the lines 
are the continuum approximation. 

\begin{figure}
	\centering
	\includegraphics[width=0.49\linewidth]{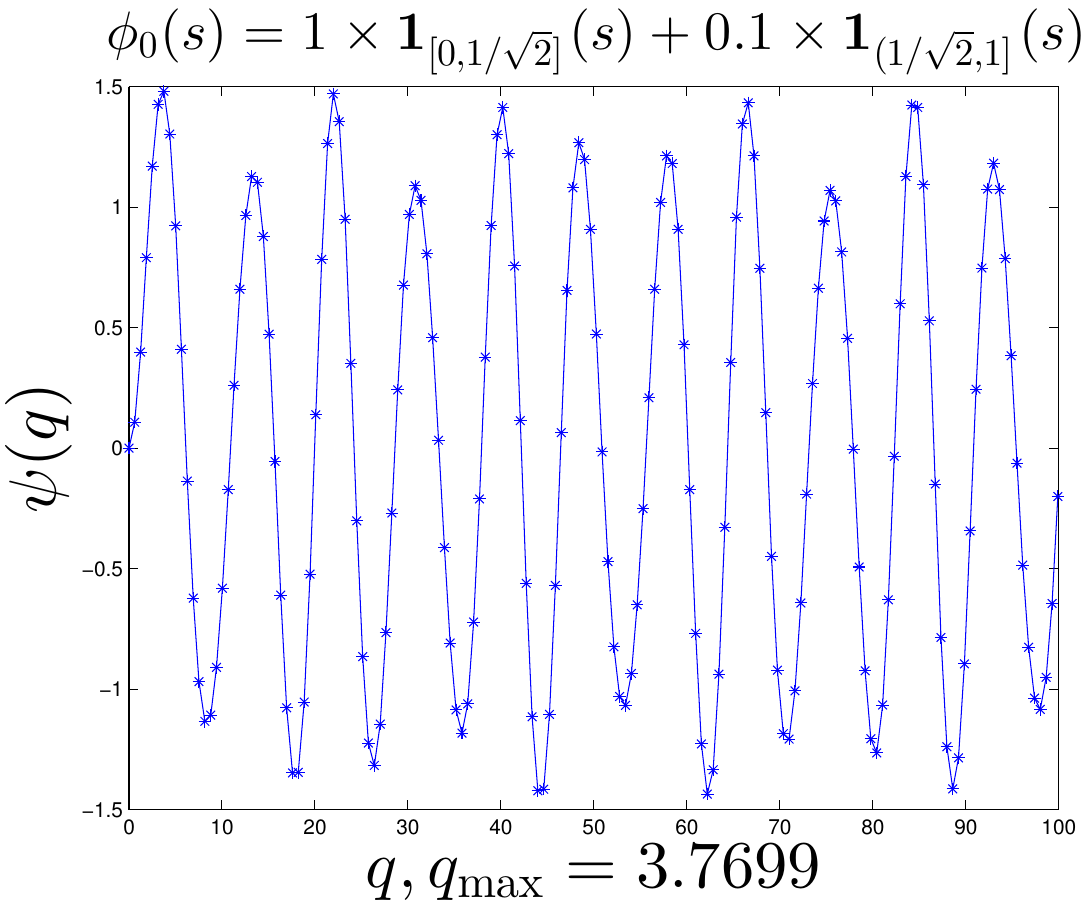}
	\includegraphics[width=0.49\linewidth]{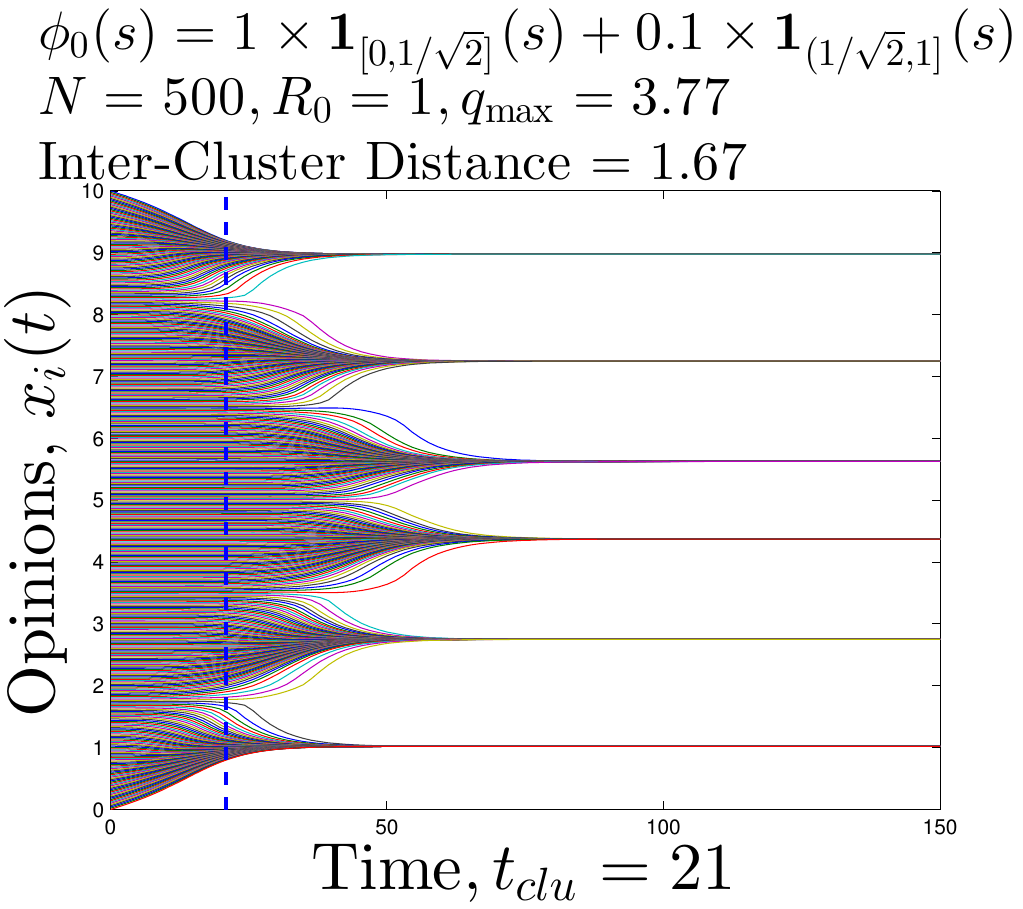}
	\caption{Simulation for $\phi_0(s)=\phi^1_0(s)$. Left: $\psi(q)$ evaluated at $R_0q\in\mathcal{K}$. 
	Right: Simulations of (\ref{eq:discrete dx_i}). 
	The vertical dashed line is at $t=t_{clu}$.}
	\label{fig:simulation of phi1}
\end{figure}

From Figure \ref{fig:simulation of phi1}, we can see that there is a unique $q_\mathrm{max}=3.7699$. From our analysis, we do not expect to have 
consensus convergence because $q_\mathrm{max}=3.7699<2\pi$. The distance between clusters is $2\pi R_0/q_\mathrm{max}=1.67$, and therefore we 
should have roughly $L/1.67=5.99$ clusters, and indeed we have 6 clusters in our simulation. In addition, 
$t_{clu}=\ln N/(2\rho_0R_0\psi(q_\mathrm{max}))=21$ (the vertical blue dashed line) also correctly predicts the time to cluster 
formation.

\begin{figure}
	\centering
	\includegraphics[width=0.49\linewidth]{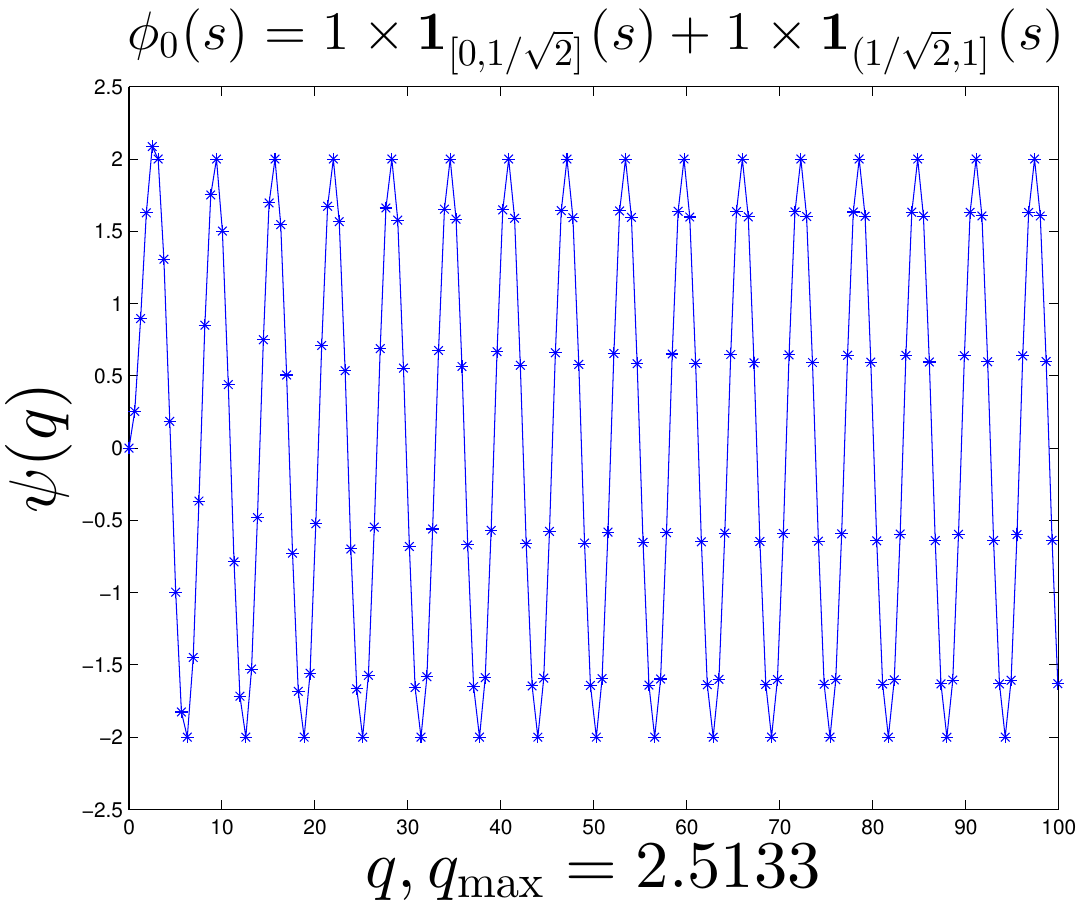}
	\includegraphics[width=0.49\linewidth]{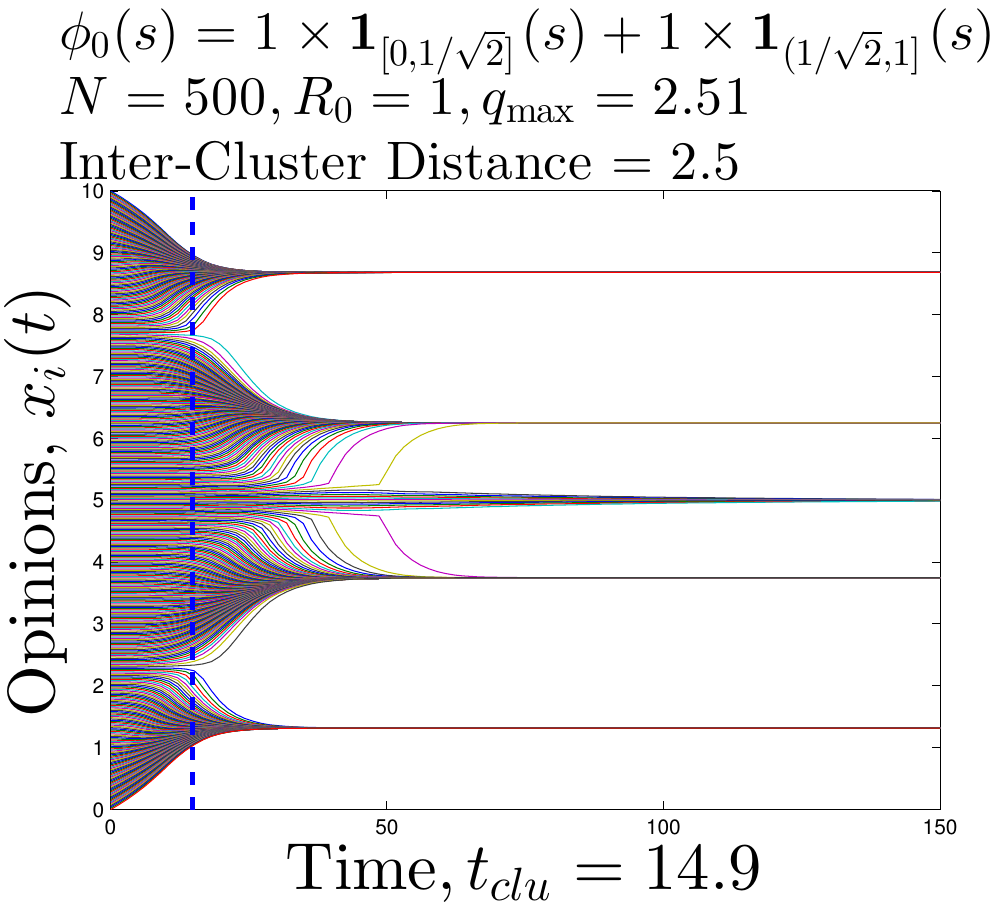}
	\caption{Simulation for $\phi_0(s)=\phi^2_0(s)$. Left: $\psi(q)$ evaluated at $R_0q\in\mathcal{K}$. 
	Right: Simulations of (\ref{eq:discrete dx_i}). 
	The vertical dashed line is at $t=t_{clu}$.}
	\label{fig:simulation of phi2}
\end{figure}

In Figure \ref{fig:simulation of phi2}, if $\phi_0(s)=\mathbf{1}_{[0,1]}(s)$, then $\psi(q)$ has a unique $q_\mathrm{max}=2.51$ but it also has 
many suboptimal $q$ where $\psi(q)$ is very close to $\psi(q_\mathrm{max})$. Note that $q_\mathrm{max}<2\pi$ means that there is no consensus convergence. 
The inter-cluster distance is $2.5$, which is correct for the top and the bottom clusters. However, the central clusters are affected by the suboptimal $q$ 
and therefore their inter-cluster distances are different. We also note that $t_{clu}=\ln N/(2\rho_0R_0\psi(q_\mathrm{max}))=14.9$ (the vertical blue 
dashed line) correctly estimates the time to the formation of the top and bottom clusters.

\begin{figure}
	\centering
	\includegraphics[width=0.49\linewidth]{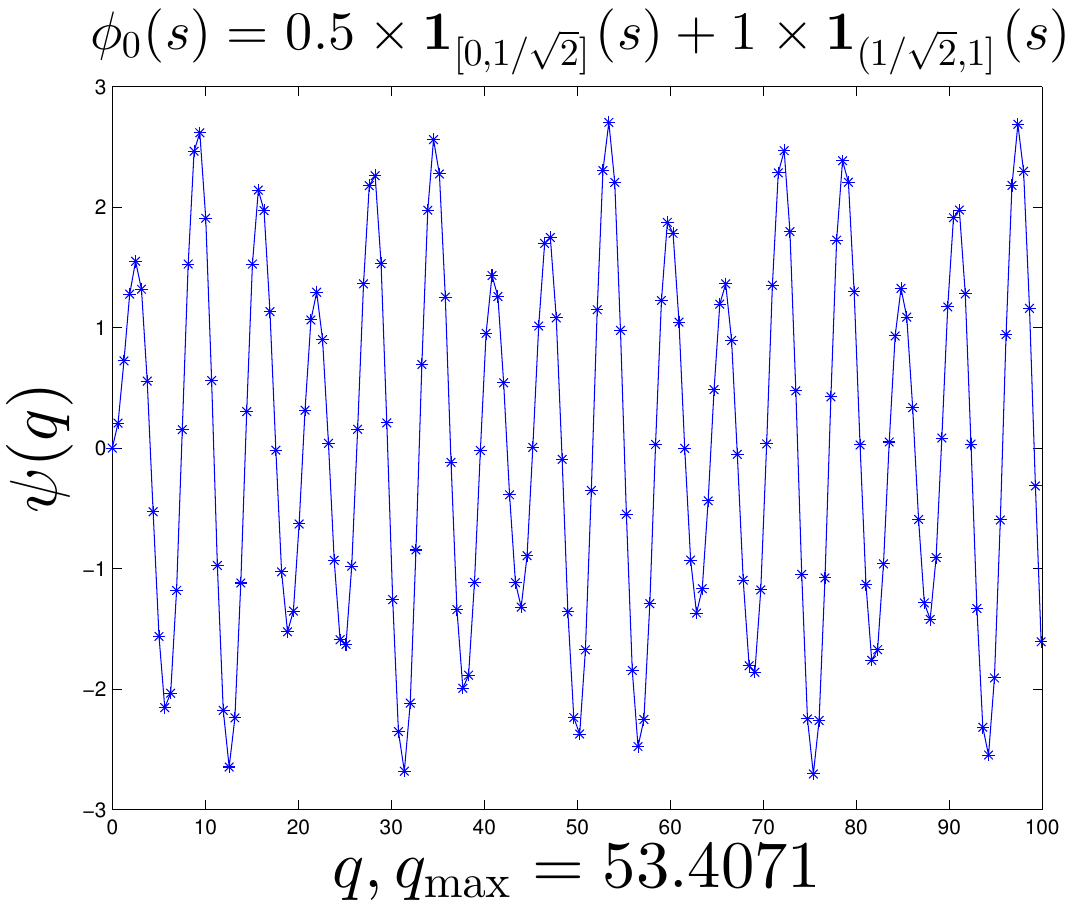}
	\includegraphics[width=0.49\linewidth]{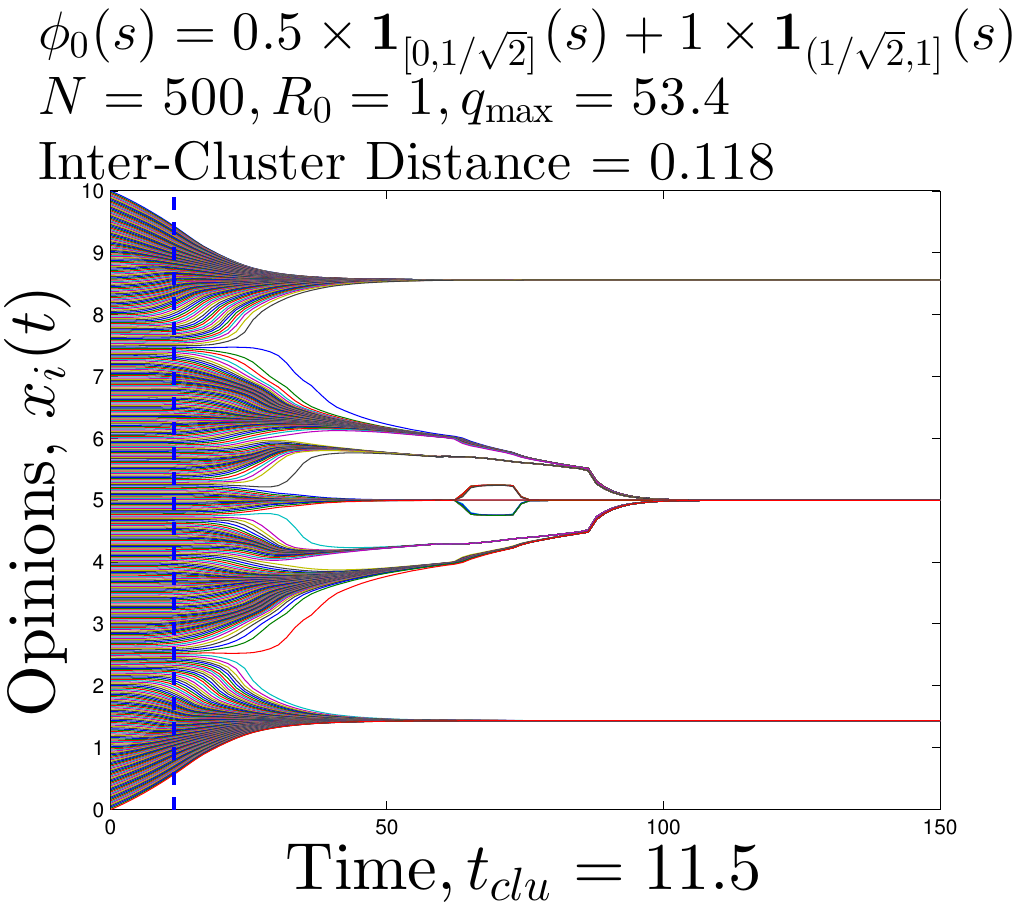}
	\caption{Simulation for $\phi_0(s)=\phi^3_0(s)$. Left: $\psi(q)$ evaluated at $R_0q\in\mathcal{K}$. 
	Right: Simulations of (\ref{eq:discrete dx_i}). 
	The vertical dashed line is at $t=t_{clu}$.}
	\label{fig:simulation of phi3}
\end{figure}

We see an interesting result in Figure \ref{fig:simulation of phi3} for 
$\phi_0(s)=0.5\times\mathbf{1}_{[0,1/\sqrt{2}]}(s) + \mathbf{1}_{(1/\sqrt{2}, 1]}(s)$. From the plot of $\psi(q)$, we can see that $q_\mathrm{max}$
might not be unique and the first few local maximizers are $q=2.5133$, $9.4248$, $15.7080, \ldots$, and the corresponding inter-cluster distances are 
$2.5$, $0.6667$, $0.4000, \ldots$. We can see from the simulation that there are two noticeable inter-cluster distances: $2.5$ and $0.6667$. For 
$R_0q\in\mathcal{K}$, $0\leq q\leq 100$, $q_\mathrm{max}=53.4071$ so that the necessary condition to have the consensus convergence 
$q_\mathrm{max}>2\pi$ is satisfied. However, we do not have consensus convergence in this case because $q_\mathrm{max}>2\pi$ is not a 
sufficient condition. We notice that although $q_\mathrm{max}$ might not be unique, $t_{clu}=11.5$ still predicts the time to cluster formation 
because it is related to $\psi(q_\mathrm{max})$ not $q_\mathrm{max}$ and thus the non-uniqueness of $q_\mathrm{max}$ does not affect $t_{clu}$.

\begin{figure}
	\centering
	\includegraphics[width=0.49\linewidth]{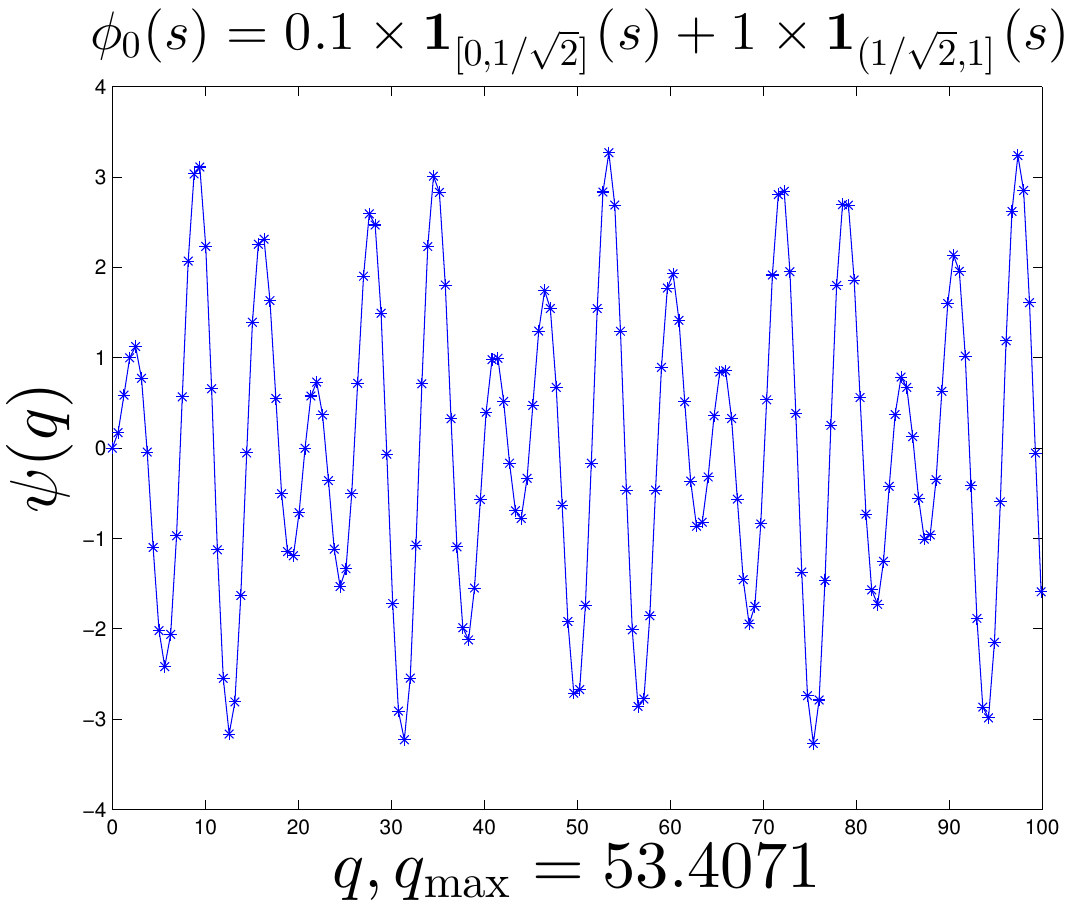}
	\includegraphics[width=0.49\linewidth]{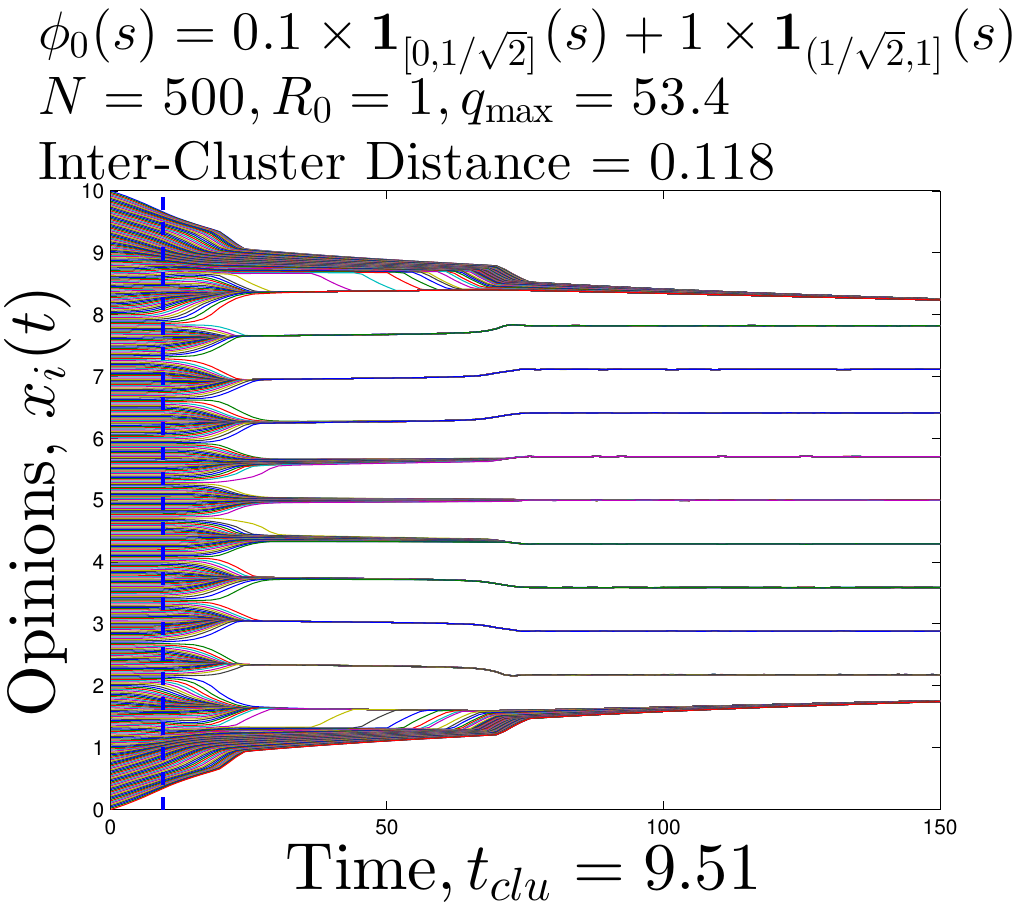}
	\caption{Simulation for $\phi_0(s)=\phi^4_0(s)$. Left: $\psi(q)$ evaluated at $R_0q\in\mathcal{K}$. 
	Right: Simulations of (\ref{eq:discrete dx_i}) for $t\leq 150$. 
	The vertical dashed line is at $t=t_{clu}$.}
	\label{fig:short simulation of phi4}
\end{figure}

\begin{figure}
	\centering
	\includegraphics[width=0.49\linewidth]{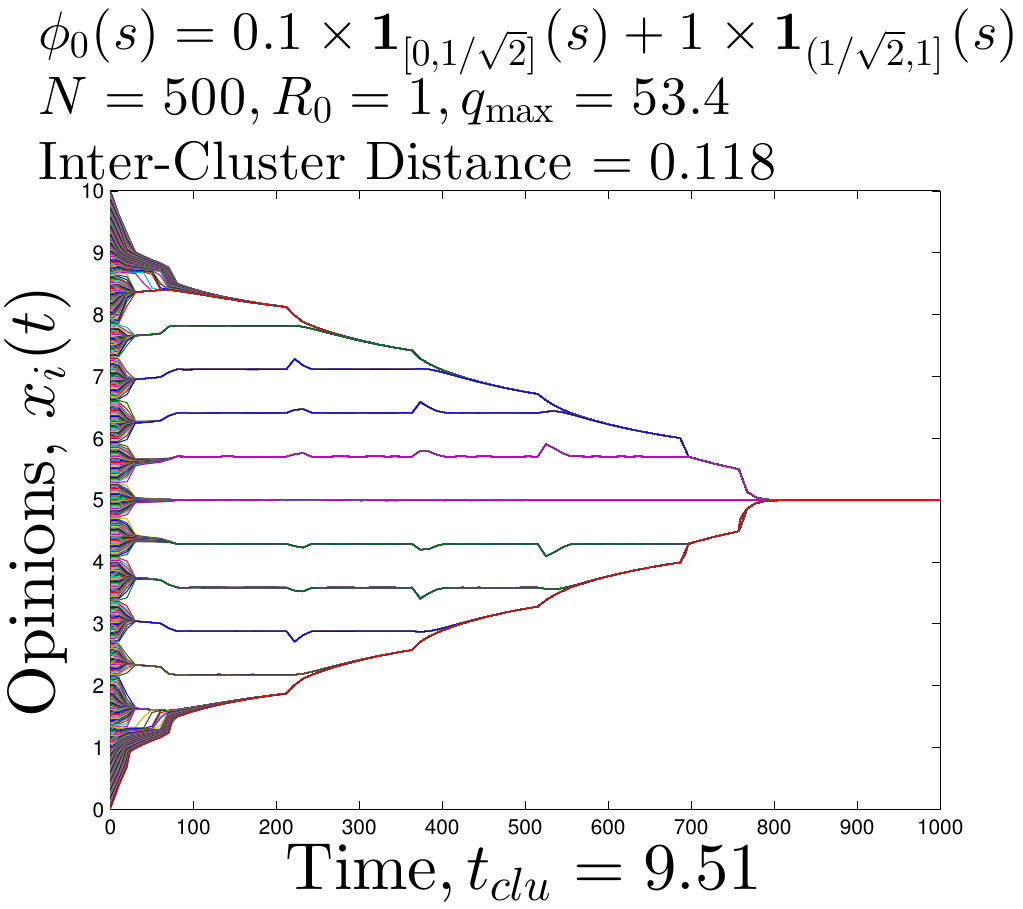}
	\caption{Simulation for $\phi_0(s)=\phi^4_0(s)$. Simulations of (\ref{eq:discrete dx_i}) for $t\leq 1000$.}
	\label{fig:long simulation of phi4}
\end{figure}

In Figure \ref{fig:short simulation of phi4} and \ref{fig:long simulation of phi4}, we see consensus convergence. From the plot of $\psi(q)$, 
we can see that $q_{\max}$ might not be unique and the first few local maximizers are $q=2.5133$, $9.4248$, $16.3363, \ldots$, and the 
corresponding inter-cluster distances are $2.5$, $0.6667$, $0.3846, \ldots$. In this case, the only noticeable inter-cluster distance is $0.6667$ and we do not 
observe the inter-cluster distance of $2.5$, because $\psi(2.5133)\ll\psi(9.4248)$. For $R_0q\in\mathcal{K}$, $0\leq q\leq 100$, $q_{\max}=53.4071$ 
so that the necessary condition to have the consensus convergence $q_{\max}>2\pi$ is satisfied and indeed we see form 
Figure \ref{fig:long simulation of phi4} that we have consensus convergence in this case.

\begin{figure}
	\centering
	\includegraphics[width=0.49\linewidth]{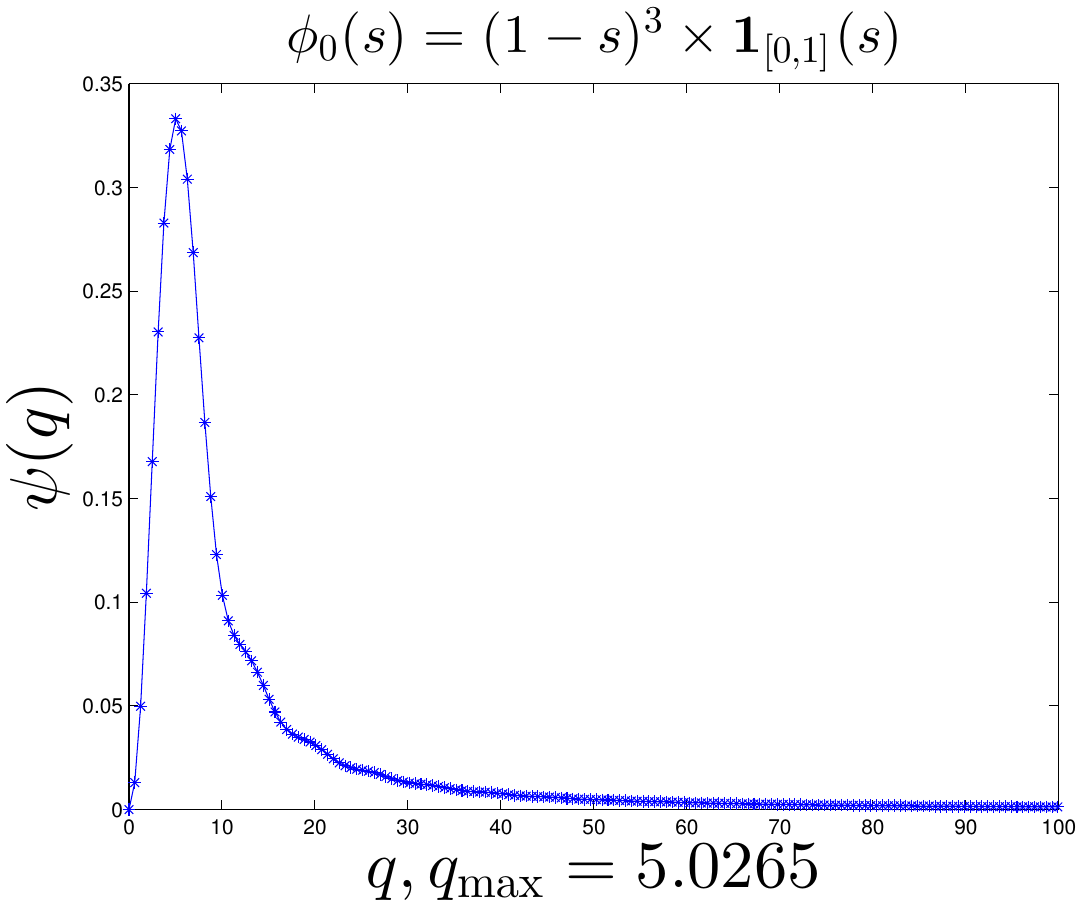}
	\includegraphics[width=0.49\linewidth]{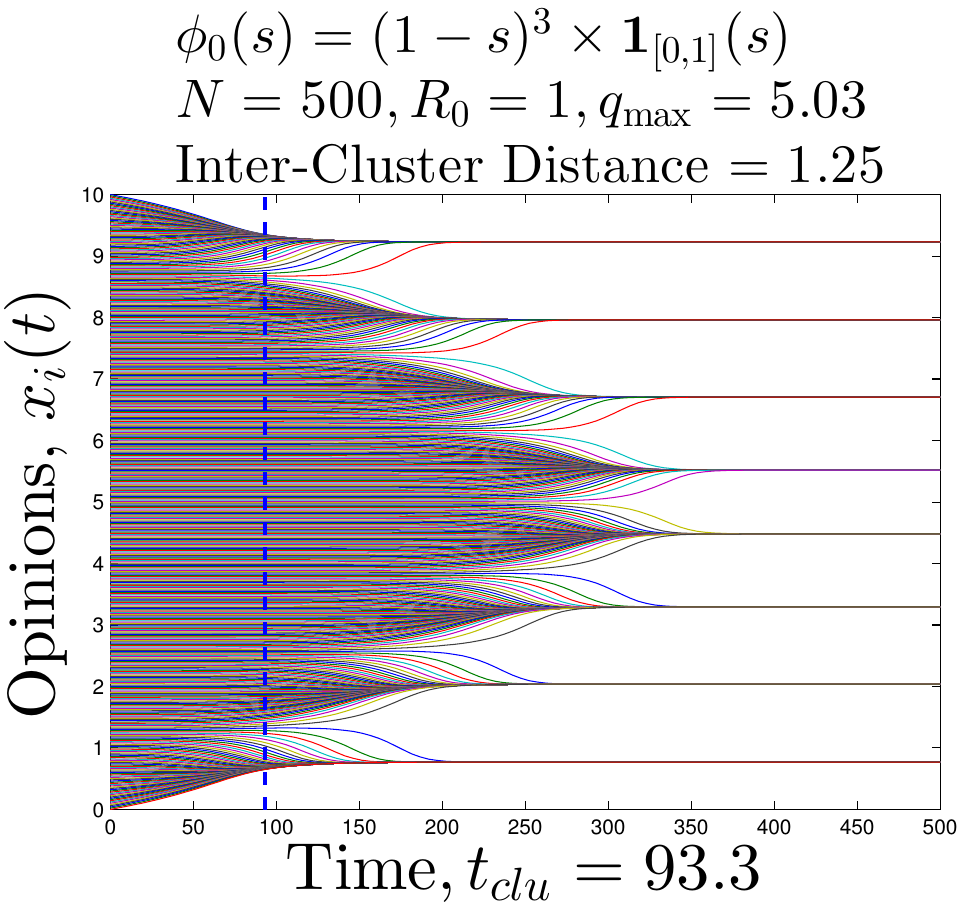}
	\caption{Simulation for $\phi_0(s)=\phi^5_0(s)$. Left: $\psi(q)$ evaluated at $R_0q\in\mathcal{K}$. 
	Right: Simulations of (\ref{eq:discrete dx_i}). The vertical dashed line is at $t=t_{clu}$.}
	\label{fig:simulation of phi5}
\end{figure}

In Figure \ref{fig:simulation of phi5}, we choose $\phi_0(s)$ so that $\psi(s)$ has a unique local maximum and $q_{\max}=5.0265$. In this case, 
$q_{\max}<2\pi$ and therefore there is no consensus convergence. The inter-distance is $1.25$ and $L/1.25=8$ which is exactly the number of 
clusters in this case. Again, $t_{clu}$ predicts the time to cluster formation very well.

\begin{figure}
	\centering
	\includegraphics[width=0.49\linewidth]{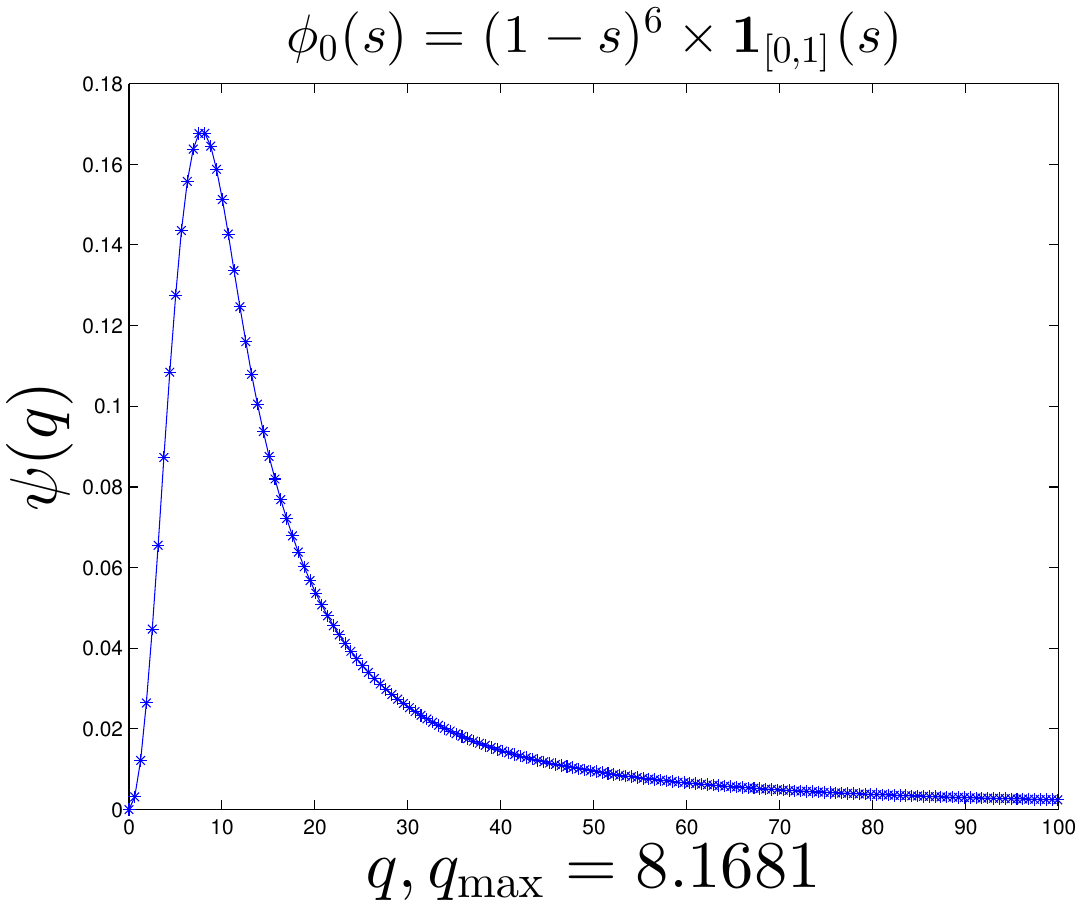}
	\includegraphics[width=0.49\linewidth]{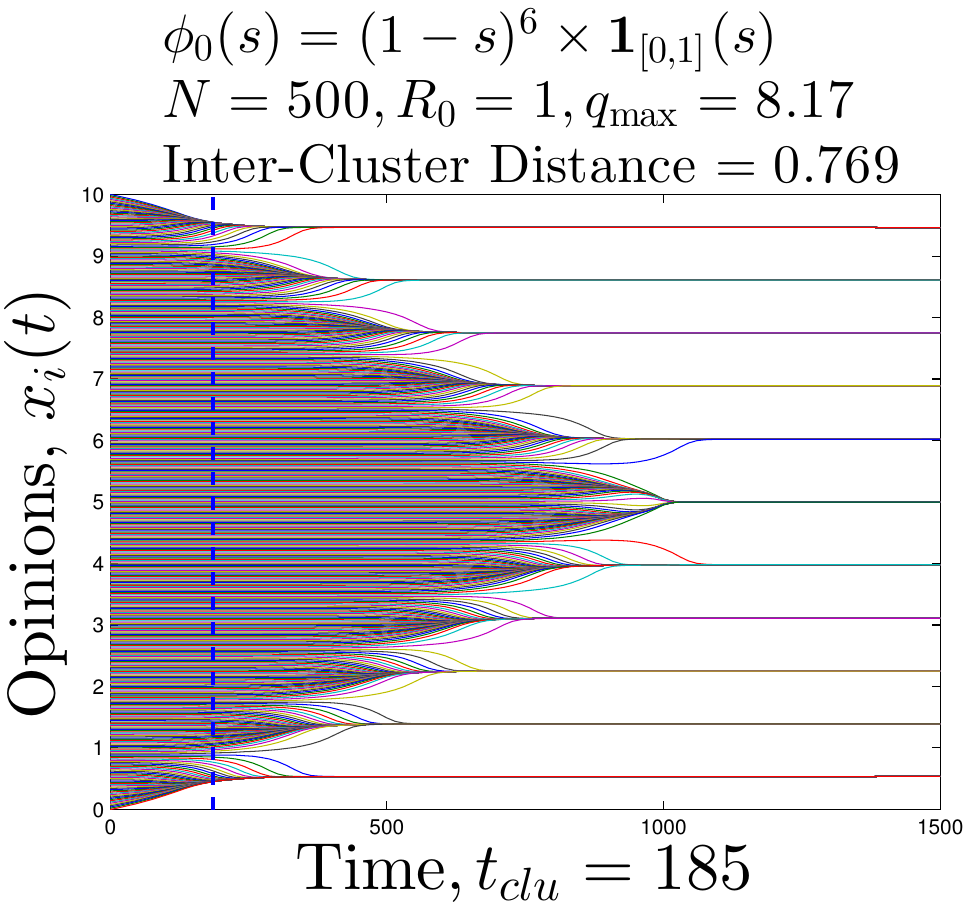}
	\caption{Simulation for $\phi_0(s)=\phi^6_0(s)$. Left: $\psi(q)$ evaluated at $R_0q\in\mathcal{K}$. 
	Right: Simulations of (\ref{eq:discrete dx_i}). The vertical dashed line is at $t=t_{clu}$.}
	\label{fig:simulation of phi6}
\end{figure}

Finally, Figure \ref{fig:simulation of phi6} considers $\phi_0(s)$ so that $\psi(s)$ has a unique local maximum, but with a larger exponent, 
$q_{\max}=8.1681>2\pi$. 
The inter-cluster distance corresponding to $q_{\max}=8.1681$ is $0.769$, which is consistent with the actual inter-cluster distance.
The quantity $L/0.769=13$ gives a good approximation for the actual number of clusters, which is $11$. As in all the previous cases, $t_{clu}=185$ 
predicts the time to cluster formation well.
Here, $q_{\max}>2\pi$, so we could expect to observe consensus convergence.
However the inter-cluster distance $0.769$ is such that $\phi(0.769)\sim 10^{-4}$, so we cannot see cluster evolution for the time horizon of the simulation.


\section{Stochastic consensus convergence: $\sigma>0$}
\label{sec:stoch}

In this section, we consider the case that $\sigma>0$ in (\ref{eq:dx_i}). In other words, the system is stochastic
and we are dealing with a nonlinear Fokker-Planck equation.

\subsection{Linear stability analysis}

As in the deterministic case, we linearize the Fokker-Planck equation (\ref{eq:Fokker-Plank equation}) with $\sigma>0$ by assuming that 
$\rho(t,x) = \rho_0 + \rho_1(t,x) = 1/L + \rho_1(t,x)$. Substituting $\rho(t,x) = \rho_0 + \rho_1(t,x)$ into (\ref{eq:Fokker-Plank equation}) and 
assuming that $\rho_1$ is a small perturbation of $\rho$ so that the $O(\rho_1^2)$ term is negligible, we find that $\rho_1(t,x)$ satisfies:

\begin{equation}
	\label{eq:linearized mean-field with positive sigma}
	\frac{\partial \rho_1}{\partial t}(t,x) 
	= \rho_0 \int \frac{\partial \rho_1}{\partial x}(t,x-y) y \phi(|y|) dy + \frac{\sigma^2}{2} \frac{\partial^2 \rho_1}{\partial x^2}(t,x).
\end{equation}
In the Fourier domain:
\begin{equation}
	\label{eq:linearized mean-field with positive sigma in Fourier domain}
	\frac{\partial \hat{\rho}_1}{\partial t}(t,k) 
	= \left[ i \rho_0 k \int e^{-iky} y \phi(|y|) dy - \frac{\sigma^2 k^2}{2} \right] \hat{\rho}_1 (t,k),
\end{equation}
which gives the growth rates of the modes:
\begin{equation}
\label{eq:stochastic gamma_k}
\gamma_k 
= \mathrm{Re} \left[ i \rho_0 k \int e^{-iky} y \phi(|y|) dy - \frac{\sigma^2 k^2}{2} \right]
= \rho_0 k \int \sin(ky) y \phi(|y|) dy - \frac{\sigma^2 k^2}{2}.
\end{equation}
We can rewrite the growth rate $\gamma_k = \rho_0 R_0 \psi_\sigma(k R_0)$, where
\begin{equation}
	\label{eq:psi_sigma}
	\psi_\sigma(q) = 2q \int_0^1 \phi_0(s) s \sin(q s) ds - \frac{\sigma^2 q^2}{2 \rho_0 R_0^3}.
\end{equation}
The optimal positive frequency is $k_{\max}$ that is the element of $\mathcal{K} = \{2\pi n /L, n \in \NN\}$ that maximizes $\psi_\sigma(k R_0)$, that is 
close to $q_{\max}/R_0$, where
\begin{equation}
	q_{\max}=\underset{q>0}{\arg\max}\left[\psi_\sigma(q)\right].
\end{equation}

There is a critical value $\sigma_c$ of $\sigma$ such that the system has a completely different overall behavior for $\sigma<\sigma_c$ and for
$\sigma>\sigma_c$. We can view $\sigma$ as the magnitude of the noise energy or temperature of the system.  There are two types of forces in the system 
(\ref{eq:dx_i}): the attractive interaction $-a_{ij} (x_i-x_j)$ and the random force $\sigma dW^i(t)$. If $\sigma<\sigma_c$, then 
the attractive interaction dominates the random force and thus the system is a perturbation of the deterministic 
system. If $\sigma>\sigma_c$, then the random force dominates, the attractive interaction is negligible, and 
therefore the overall system behaves like a system of $N$ independent random processes.

The above observations can be articulated mathematically. Let
\begin{align*}
	\sigma_c^2 
	&= \max_{q>0} \left[\frac{4\rho_0 R_0^3}{q}\int_0^1 \phi_0(s) s \sin(qs) ds\right]\\
	&= \max_{q>0} \left[4\rho_0 R_0^3\int_0^1 \phi_0(s) s^2 \frac{\sin(qs)}{qs} ds\right]
	=4 \rho_0 R_0^3 \int_0^1 s^2 \phi_0(s)ds.
\end{align*}
If $\sigma<\sigma_c$, then from (\ref{eq:psi_sigma}) we find that 
\begin{equation}
	\max_{q>0}\psi_\sigma(q) > 0,
\end{equation}
and $\hat{\rho}(t,k_{\max})$ has positive growth rate $\gamma_{\max} = \rho_0 R_0 \psi_\sigma( k_{\max} R_0)$. The linear system 
(\ref{eq:linearized mean-field with positive sigma in Fourier domain}) is unstable, which is analogous to the deterministic case.

If $\sigma>\sigma_c$, then all of $\hat{\rho}(t,k)$ have negative growth rates. In other words, the constant density is linearly 
stable and therefore the overall system is stable, since this is what linear stability implies in this case.

We note that the same technique for computing $\sigma_c$, with linear stability analysis for different noisy opinion models, is also used in 
\cite{Pineda2009}\cite{Pineda2011}\cite{Pineda2013}.

\subsection{Fluctuation theory}

Since our goal is to analyze the behavior of clusters, we suppose from now on that $\sigma<\sigma_c$. 

The fluctuation analysis of the stochastic model is similar to that of the deterministic case. 
If $x_1(0), \ldots, x_N(0)$ are independent, uniform random variables in $[0,L]$, then 
\begin{equation*}
	\rho^N_1(t=0,dx) :=  \sqrt{N} \left( \rho^N(t=0,dx) - \rho_0(dx) \right)
	= \sqrt{N} \left(\frac{1}{N} \sum_{j=1}^{N}\delta_{x_j(0)}(dx) - \frac{dx}{L} \right)
\end{equation*}
converges in distribution as $N \to \infty$ to the measure $\rho_1(t=0,dx)$, whose frequency components
\begin{equation*}
	\hat{\rho}_1(t=0,k) = \lim_{N\to\infty} \sqrt{N} \left( \frac{1}{N}\sum_{j=1}^{N} e^{-ikx_j(0)}\right)
\end{equation*}
are independent complex circular Gaussian random variables, with mean zero and variance $1$ for $k \in {\cal K}\backslash \{0\}$:
\begin{equation*}
	\EE \left[ \hat{\rho}_1(t=0,k) \right]=0,\quad
	\EE \left[ \hat{\rho}_1(t=0,k) \overline{\hat{\rho}_1(t=0,k')} \right] = \delta_{kk'},\quad 
	k,k' \in {\cal K} \backslash \{0\},
\end{equation*}
$\hat{\rho}_1(t=0,-k) = \overline{ \hat{\rho}_1(t=0,k) }$, while $\hat{\rho}_1(t=0,k=0) =0$.

For any $T<\infty$, the measure-valued process 
\begin{equation}
	\rho^N_1(t,dx) :=  \sqrt{N} \left( \rho^N(t,dx) - \rho_0(dx) \right) , \quad t \in [0,T]
\end{equation}
converges in distribution as $N \to \infty$ to a measure-valued process $\rho_1(t,dx)$ whose density $\rho_1(t,x)$ satisfies a stochastic PDE 
(see \cite{Dawson1983}):
\begin{equation}
	\label{eq:stochastic rho1}
	d\rho_1 (t,x) 
	= \left[\rho_0 \int \frac{\partial \rho_1}{\partial x}(t,x-y) y \phi(|y|) dy  + \frac{\sigma^2}{2} \frac{\partial^2 \rho_1}{\partial x^2} (t,x)\right] dt
	+ \sigma d W(t,x)
\end{equation}
with the random initial condition described above. Here $W(t,x)$ is a space-time Gaussian random noise with mean zero and covariance
\begin{equation}
	\label{eq:covariance of Gaussian noise}
	{\rm Cov}\left(\int_0^L W(s,x) f_1(x) dx, \overline{\int_0^L W(t,x) f_2(x) dx}\right) = \frac{\min\{s,t\}}{L} \int_0^L f'_1(x) \overline{f'_2(x)}   dx
\end{equation}
for any test functions $f_1(x)$ and $f_2(x)$.  The Fourier transform of $W(t,x)$ is $\hat{W}(t,k) = \int_0^L W(t,x) e^{-ikx} dx$ for $k\in {\cal K}$.  
From (\ref{eq:covariance of Gaussian noise}), we see that $\{\hat{W}(t,k), k\in {\cal K}, k \geq 0\}$ are 
independent, complex-valued Brownian motions with the variance: 
\begin{equation}
	\label{eq:variance of Fourier transform of Gaussian noise}
	{\rm Cov} \left(\hat{W}(t,k), \overline{\hat{W}(t,k)}\right) = \frac{t}{L} \int_0^L (-ik)e^{-ikx} \overline{(-ik)e^{-ikx}} dx = t k^2,
\end{equation}
and $\hat{W}(t,-k)= \overline{\hat{W}(t,k)}$.
Taking the Fourier transform on (\ref{eq:stochastic rho1}), for each $k\in {\cal K}$, $\hat{\rho}_1(t,k) = \int_0^L \rho_1(t,x) e^{-ikx} dx$ 
is a complex-valued Ornstein-Uhlenbeck (OU) process:
\begin{equation}
	\label{eq:complex OU}
	d\hat{\rho}_1 (t,k) 
	= \left[ i \rho_0 k \int_0^L e^{-iky} y \phi(|y|) dy - \frac{\sigma^2 k^2}{2} \right] \hat{\rho}_1 (t,k) dt
	+ \frac{\sigma k}{\sqrt{2}} d ( W^{(k)}(t) + i \tilde{W}^{(k)}(t)).
\end{equation}
Here $\{W^{(k)}(t), k\in {\cal K}, k>0\}$ and $\{\tilde{W}^{(k)}(t), k\in {\cal K}, k>0\}$ are independent real Brownian motions,
$W^{(0)}(t)=\tilde{W}^{(0)}(t)=0$, ${W}^{(-k)}(t)=-{W}^{(k)}(t)$, and $\tilde{W}^{(-k)}(t)=\tilde{W}^{(k)}(t)$.
The equation (\ref{eq:complex OU}) is solvable and we have, for any $k \in {\cal K}\backslash \{0\}$,
\begin{align}
	\label{eq:complex rho_1}
	\hat{\rho}_1 (t,k) &= e^{\alpha_k t} \hat{\rho}_1 (0,k) + \frac{\sigma k}{\sqrt{2}} \int_0^t e^{\alpha_k (t-s)} d ( W^{(k)}(s)+ i \tilde{W}^{(k)}(s)) ,\\
	\alpha_k &= i \rho_0 k \int_0^L e^{-iky} y \phi(|y|) dy - \frac{\sigma^2 k^2}{2}. \notag
\end{align}
In particular, $\hat{\rho}_1(t,k=0)=0$. Because $\{\hat{\rho}_1 (0,k), W^{(k)}(t), \tilde{W}^{(k)}(t), k\in {\cal K}, k>0\}$ are independent,  $\{\hat{\rho}_1 (t,k), k\in {\cal K}, k>0\}$ are 
independent OU processes with mean zero and variance 
\begin{equation}
	\label{eq:variance of rho_1}
	\EE \left[ \hat{\rho}_1(t,k) \overline{\hat{\rho}_1(t,k)} \right] 
	= e^{2 \gamma_k t} + \sigma^2 k^2 \int_0^t e^{2\gamma_k (t-s)} ds
	= e^{2 \gamma_k t}\left[1 + \frac{\sigma^2 k^2}{2\gamma_k}\right],
\end{equation}
for $k\in {\cal K}$, $k>0$, where $\gamma_k$ is the real part of $\alpha_k$. In addition, because $\alpha_{-k} = \overline{\alpha_k}$ and 
$W^{(-k)}(t)=-W^{(k)}(t)$, we have $\hat{\rho}_1(t,-k) = \overline{\hat{\rho}_1(t,k)}$ for $k\in {\cal K}\backslash\{0\}$. Finally, $\hat{\rho}_1(t,0) = 0$.
Therefore, 
\begin{align*}
	&\EE\left[\rho_1(t,x)\rho_1(t,x')\right]
	= \EE\left[\sum_k\hat{\rho}_1(t,k)\frac{e^{ikx}}{L} \sum_k\hat{\rho}_1(t,k)\frac{e^{ikx'}}{L}\right]\\
	&\quad = \sum_k \EE\left[\hat{\rho}_1(t,k)\frac{e^{ikx}}{L} \hat{\rho}_1(t,-k)\frac{e^{-ikx'}}{L}\right]
	= \frac{1}{L^2} \sum_{k\neq 0} e^{2 \gamma_k t}e^{ik(x-x')}\left[1 + \frac{\sigma^2 k^2}{2\gamma_k}\right].
\end{align*}

As $t$ increases, the spectrum of $\rho_1(t,x)$ becomes concentrated around the optimal wavenumber $k_{\max}$.  In addition, we note that $k^2/\gamma_k$ is bounded 
and the term $\sigma^2 k^2/\gamma_k$ is negligible if $\sigma$ is sufficiently small. We can assume $\sigma$ is small because we need $\sigma<\sigma_c$ for cluster 
formation.  If $\sigma^2 k^2/\gamma_k$ is negligible and $L^2 \gg 4|\gamma_{\max}''| t$, we can expand $\gamma_k = \gamma_{\max} + \frac{1}{2} \gamma_{\max}'' (k-k_{\max})^2$ for $k$ around $k_{\max}$, and use a continuum approximation for the discrete sum as we do in the deterministic case:
\begin{equation*}
	\EE\left[\rho_1(t,x)\rho_1(t,x')\right] 
	\simeq \left(\frac{1}{L}e^{2\gamma_{\max}t}\cos ( k_{\max}(x-x')) \right)
	\left(\frac{1}{\sqrt{\pi|\gamma_{\max}''| t}} e^{-\frac{(x-x')^2}{4|\gamma_{\max}''| t}}\right).
\end{equation*}
If $L^2 \ll 4|\gamma_{\max}''| t$, then the continuum approximation is not valid and in this case
\begin{equation*}
	\EE \left[ \rho_1(t,x) \rho_1(t,x') \right] \simeq \frac{2}{L^2} e^{ 2 \gamma_{\max} t}  
	\cos \left(k_{\max} (x-x')\right) \left[1 + \frac{\sigma^2 k_{\max}^2}{2\gamma_{\max}}\right].
\end{equation*}

Because $\sigma<\sigma_c$ we have that $\gamma_{\max}>0$ and then the linear system (\ref{eq:stochastic rho1}) is unstable and therefore the central limit 
theorem breaks down when $\rho_1(t,x)/\sqrt{N}$ is no longer smaller than $\rho_0=1/L$. More precisely, the time $t_{clu}$ for the onset of 
clustering is when $\EE \big[ \rho_1(t_{\rm clu},x)^2 \big] \simeq N L^{-2}$, which is approximately
\begin{equation*}
	t_{clu} \simeq \frac{1}{2 \gamma_{\max} } \ln N \simeq \frac{1}{2 \rho_0 R_0  \psi_\sigma(q_{\max} )} \ln N 
\end{equation*}
when $N \gg 1$.

\subsection{Consensus convergence}

We assume $\sigma <\sigma_c$ so that there are unstable modes for the linearized evolution, 
which means that there is clustering. The number of and the distance between clusters can 
be estimated with $q_{\max}$. We find that the first term of the right side of (\ref{eq:psi_sigma}) is bounded while the second term of 
(\ref{eq:psi_sigma}) is quadratic with negative leading coefficient. Therefore, increasing $\sigma$ tends to reduce $q_{\max}$, that is to say, 
to increase the mean distance between clusters.

Let us consider the case that $q_{\max} < 2\pi$. From the analysis of the deterministic case, the system initially has no consensus 
convergence and there are several clusters. After clustering, the clusters do not interact with each other, but their centers move like independent Brownian 
motions. When two clusters come close to each other, within a distance $R_0$, they interact and merge. Therefore, we will eventually have 
consensus convergence, because two Brownian motions always collide in $\RR$. This can be extended to the multi-dimensional case, but 
then the conclusion can be different: in high dimension two Brownian motions may not collide. However, with periodic boundary conditions, two 
Brownian motions will always come close to each other, within a distance $R_0$.

When $q_{\max} < 2\pi$ and $\sigma$ is small then there are several clusters, after the cluster formation time. The fraction of 
agents in a cluster is the agent density times the inter-distance of the clusters:
\begin{equation*}
	m_c = \rho_0\frac{ 2\pi R_0}{q_{\max}} = \rho_0\frac{ 2\pi }{k_{\max}}.
\end{equation*}
Then the $j$-th cluster consists of about $N m_c$ agents. We assume that $\sigma$ is small enough so that 
$\sigma^2 k_{\max} \ll 2 \pi \rho_0 \phi_0(0) R_0^2$. By using the fact that the agents in a cluster stay close to each other, we can replace 
$\phi(x_i-x_j)$ by $\phi(0)$, and the agents in the $j$-th cluster have the approximate dynamics:
\begin{equation*}
	dx^{(j)}_i = -\frac{\phi_0(0)}{N} \sum_{l=1}^{Nm_c}(x^{(j)}_i-x^{(j)}_l) dt+ \sigma dW^{(j,i)}(t).
\end{equation*}
The center $X^{(j)}(t)=\frac{1}{Nm_c}\sum_{i=1}^{Nm_c}x^{(j)}_i(t)$ satisfies:
\begin{equation}
	\label{eq:center of cluster}
	X^{(j)}(t) = X^{(j)}(0) +   \frac{\sigma}{\sqrt{Nm_c}} W^{(j)}(t),
\end{equation}
where $\{W^{(j)}(t)\}$ are independent standard Brownian motions.

When $N$ is large, the empirical density $\frac{1}{Nm_c}\sum_{i=1}^{Nm_c}\delta_{x^{(j)}_i}(dx)$ is approximately a Gaussian density
\begin{equation}
	\label{eq:density of each cluster}
	\rho^{(j)} (t,dx) = \frac{1}{\sqrt{\pi(\sigma^{(j)})^2}} \exp \left( - \frac{(x-X^{(j)}(t))^2}{(\sigma^{(j)})^2}\right)dx, \quad
	\sigma^{(j)} = \frac{\sigma}{\sqrt{ m_c \phi_0(0)}}.
\end{equation}
For this argument to be valid, we must have that $\sigma^{(j)}$, the width of $\rho^{(j)}$, is much smaller than $R_0$, which 
is equivalent to our assumption $\sigma^2 k_{\max} \ll 2 \pi \rho_0 \phi_0(0) R_0^2$.

This cluster dynamics is valid as long as the centers $\{X^{(j)}(t)\}$ stay away from each other by a distance larger than $R_0$. The clusters move, 
according to independent Brownian motions with quadratic variation $\frac{\sigma^2 t}{Nm_c}$. When two clusters come close to each other within a 
distance $R_0$, they merge. Indeed, once the two centers are within distance $R_0$, they obey the following differential equations to leading order 
in $N$:
\begin{align*}
	\frac{d X^{(k)}(t)}{dt}  &= - m_c ( X^{(k)}(t) - X^{(l)}(t) ) \phi( X^{(k)}(t) - X^{(l)}(t) ),\\
	\frac{d X^{(l)}(t)}{dt}  &= - m_c ( X^{(l)}(t) - X^{(k)}(t) ) \phi( X^{(k)}(t) - X^{(l)}(t) ),
\end{align*}
which shows that the inter-cluster distance converges exponentially fast to zero and the center converges to the average of the centers just before 
collision. The number of agents or mass of the new cluster is the sum of the masses of the two clusters, the inverse square width of its empirical density is the sum of the 
inverse squares of the two widths, its center is at the weighted average  (weighted by the masses) of the two centers just before collision and it moves as 
a Brownian motion whose diffusion constant is defined in terms of its new mass. Then the cluster centers move according to Brownian motions until 
two of them get within the distance $R_0$ from each other and a new merge event occurs. This eventually forms a Markovian dynamics described in the 
next section.

\subsection{Markovian dynamics of the clusters}

After the initial clusters are formed, we can use an iterative argument to mathematically describe how all the opinions converge eventually. In the 
initial configuration, at time $\tau_0$, there are $M(\tau_0) = L k_{\max} / 2\pi $ clusters with centers 
$X^{(j)}(\tau_0) =  j 2 \pi /k_{\max}$ (up to a global shift), widths $\sigma^{(j)}(\tau_0)=\sigma/\sqrt{m_c \phi_0(0)}$, and masses 
$m^{(j)}(\tau_0) = m_c$ for $j=1,\ldots,M(\tau_0)$.

For $t\geq \tau_{n-1}$, there are $M(\tau_{n-1})$ clusters moving as 
\begin{equation*}
	X^{(j)}(t) = X^{(j)}(\tau_{n-1}) + \sigma^{(j)}(\tau_{n-1})  \frac{\sqrt{\phi_0(0)}}{\sqrt{N}} \left( W^{(j)}(t) - W^{(j)}(\tau_{n-1}) \right)	
\end{equation*}
until the stopping time
\begin{equation*}
	\tau_n = \inf \left\{ t > \tau_{n-1}: |X^{(k)}(t) -X^{(l)}(t) | = R_0, \mbox{ for some } k\neq l\right\}.
\end{equation*}
Then the two colliding clusters (with indices $k$ and $l$) merge with the new center 
\begin{equation*}
	\tilde{X}(\tau_n^+) = \frac{m^{(k)}(\tau_{n-1}) X^{(k)}(\tau_{n}^-)	+ m^{(l)}(\tau_{n-1}) X^{(l)}(\tau_{n}^-) }
	{ m^{(k)}(\tau_{n-1}) + m^{(l)}(\tau_{n-1}) },
\end{equation*}
the new mass
\begin{equation*}
	\tilde{m}(\tau_n) =  m^{(k)}(\tau_{n-1}) + m^{(l)}(\tau_{n-1})
\end{equation*}
and the new width
\begin{equation*}
	\tilde{\sigma} (\tau_n ) = \frac{\sigma}{\sqrt{\tilde{m}(\tau_n) \phi_0(0)}}
\end{equation*}
The clusters are relabeled to take into account this merging so that there are $M(\tau_n)=M(\tau_{n-1})-1$ clusters. The above process is repeated 
until $n=(L k_{\max} / 2\pi)-1$, when we have only one cluster, and hence consensus convergence.

Note that the time scale at which collisions and merges occur is of the order of $N$, as the Brownian motions are scaled by $1/\sqrt{N}$.

\subsection{Numerical simulations}

We use the explicit Euler scheme to simulate the stochastic opinion dynamics (\ref{eq:dx_i}) when $\sigma>0$:
\begin{equation}
	\label{eq:stochastic discrete dx_i}
	x_i^{n+1} = x_i^{n} -\frac{1}{N} \sum_{j=1}^N \phi(|x_i^n-x_j^n|) (x_i^n-x_j^n) \Delta t + \sigma \Delta W^{n+1}_i,\quad \phi(s) = \phi_0(s/R_0),
\end{equation}
where $\{W^{n+1}_i\}$ are independent Gaussian random variables with mean zero and variance $\Delta t$.

Our analysis is on the torus $[0,L]$, but we simulate (\ref{eq:stochastic discrete dx_i}) on $[0,L]$ with reflecting boundary conditions. As we will 
see, the simulation results agree with the analysis under periodic assumption. Because here we focus on the effects of the randomness, for 
simplicity we will work only on the case that $\phi_0(s)=\mathbf{1}_{[0,1]}(s)$.

We compute the key quantity $q_{\max}$ by exploring all possible $q$ in $[0,100]$:
\begin{equation*}
	q_{\max} = \underset{R_0q\in\mathcal{K}, 0<q\leq 100}{\arg\max} 
	\left[2q \int_0^1 \phi_0(s) s \sin(q s) ds - \frac{\sigma^2 q^2}{2 \rho_0 R_0^3}\right].
\end{equation*}
We see form the plots of $\psi_\sigma(s)$ that the randomness reduces the possibility of the non-uniqueness of $q_{\max}$ because it adds 
a negative quadratic term in $\psi_\sigma(s)$. With randomness, all of our test cases have a clear, unique $q_{\max}$. 

The parameters we use for the simulation are $\Delta t=0.1$, $L=10$, $R_0=1$ and $N=500$. For each $\sigma$, we also plot the function 
$\psi_\sigma(s)$ in (\ref{eq:psi_sigma}); the stars in the plots are the values of $\psi(s)$ evaluated at $R_0q\in\mathcal{K}$ and the lines 
are the continuum approximation.

We first to test for the effect of $\sigma_c$, the critical value for $\sigma$, which makes the system stable or unstable. In our setting, 
$\sigma_c=4 \rho_0 R_0^3 \int_0^1 s^2 \phi_0(s)ds = 0.365$ and we simulate (\ref{eq:stochastic discrete dx_i}) for $\sigma=0.1, 0.2, 0.365, 0.5$ 
that are values below, equal to and above $\sigma_c$, respectively.

\begin{figure}
	\centering
	\includegraphics[width=0.49\linewidth]{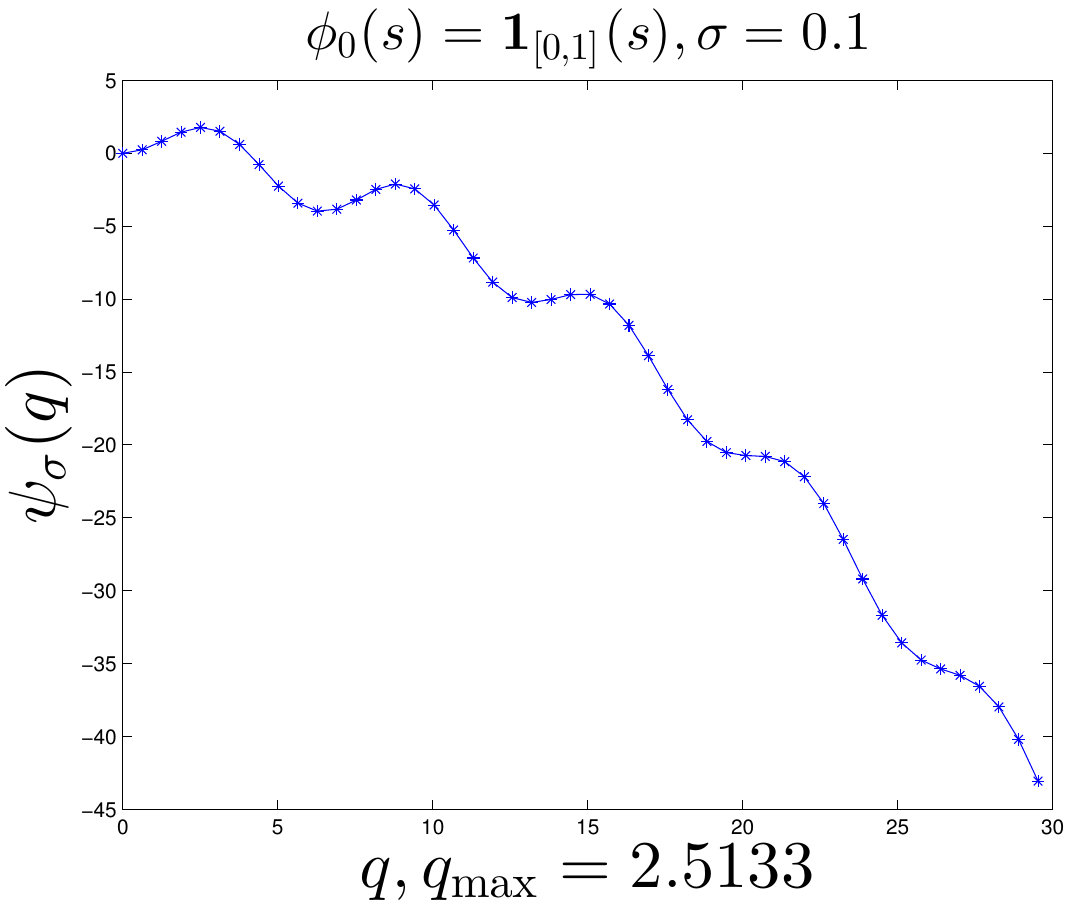}
	\includegraphics[width=0.49\linewidth]{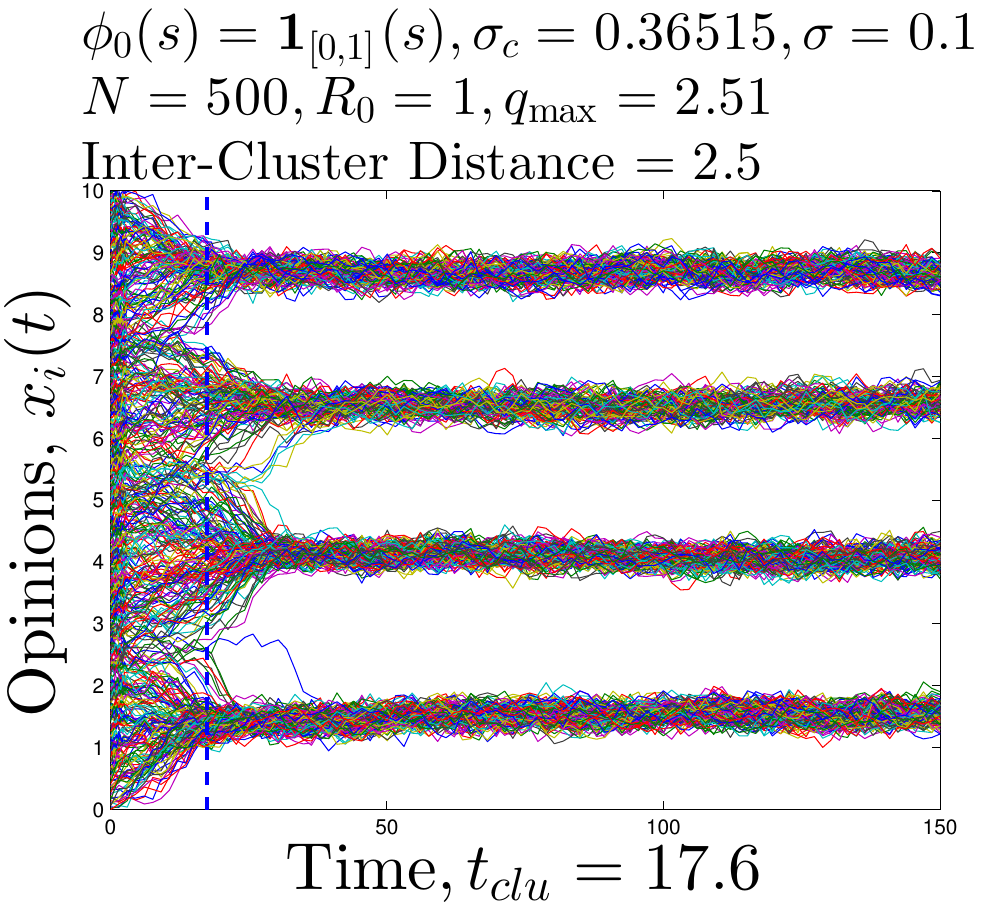}
	\caption{Simulations for $\sigma=0.1$. Left: $\psi_\sigma(q)$ evaluated at $R_0q\in\mathcal{K}$. 
	Right: Simulations of (\ref{eq:stochastic discrete dx_i}). The vertical dashed line is at $t=t_{clu}$.}
	\label{fig:sigma_c for sigma0.1}
\end{figure}

From Figure \ref{fig:sigma_c for sigma0.1}, we see that $\psi_\sigma(q)$ decreases quadratically and has the unique maximum at $q=2.5133$. However, 
$\max_{R_0q\in \mathcal{K}}\psi_\sigma(q)$ is still positive so the linearized system 
(\ref{eq:linearized mean-field with positive sigma in Fourier domain}) is still unstable. Therefore, the overall system behavior is similar to the 
deterministic case and can be viewed as a perturbed non-random opinion dynamics.

\begin{figure}
	\centering
	\includegraphics[width=0.49\linewidth]{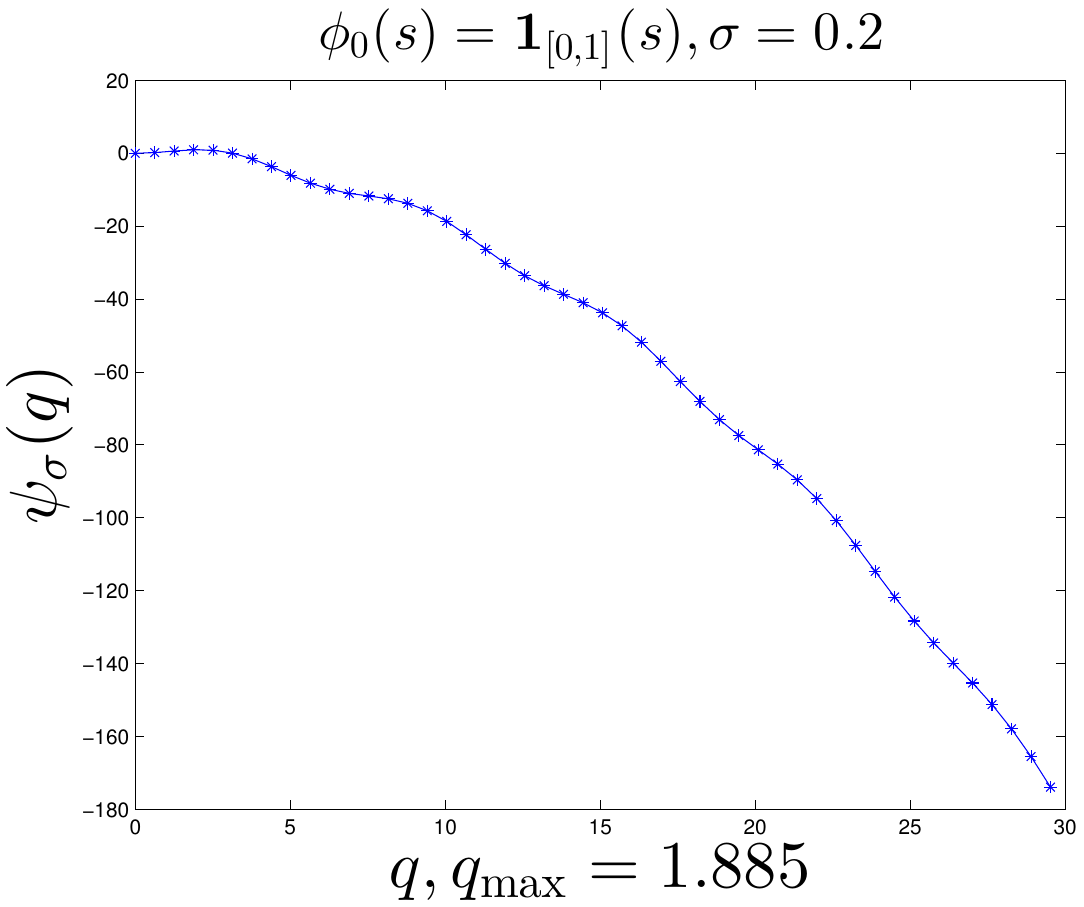}
	\includegraphics[width=0.49\linewidth]{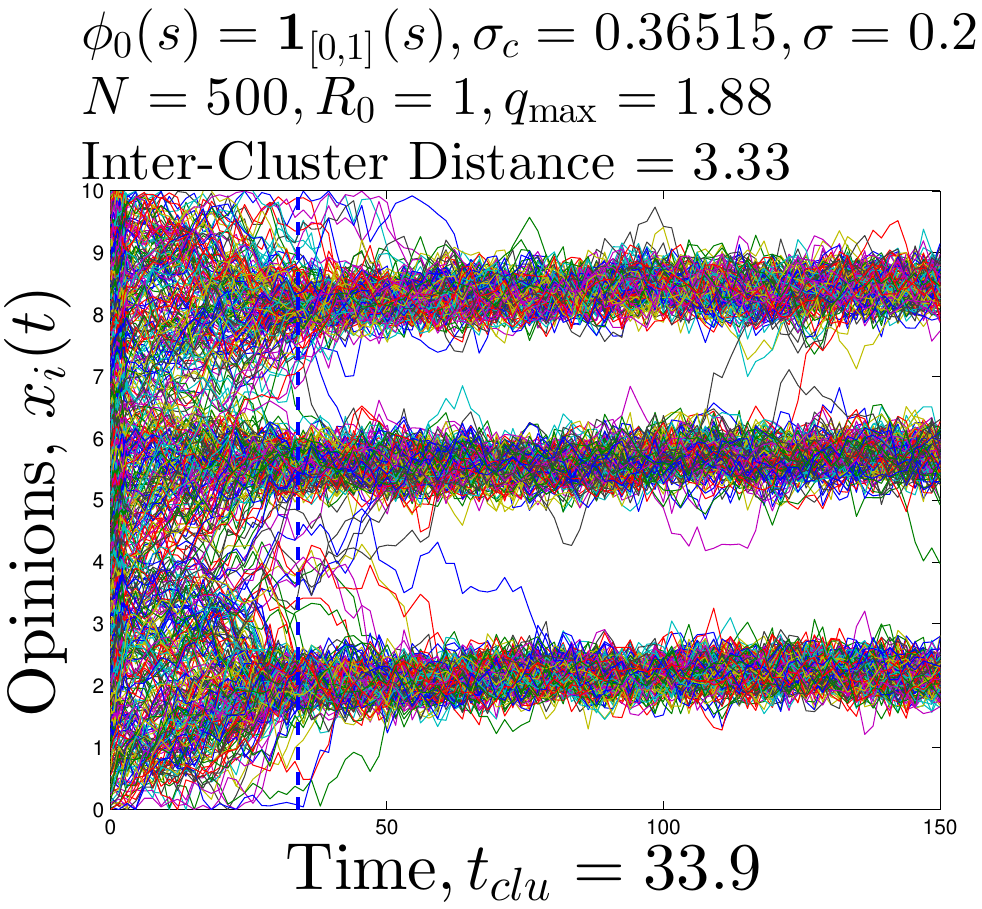}
	\caption{Simulations for $\sigma=0.2$. Left: $\psi_\sigma(q)$ evaluated at $R_0q\in\mathcal{K}$. 
	Right: Simulations of (\ref{eq:stochastic discrete dx_i}).}
	\label{fig:sigma_c for sigma0.2}
\end{figure}

We increase $\sigma$ by setting $\sigma=0.2$ and the result is in Figure \ref{fig:sigma_c for sigma0.2}. We see that as $\sigma$ increases, 
the random noise starts to affect the overall system, and the width and the inter-cluster distances become larger so we observe fewer 
clusters. Since $\max_{R_0q\in \mathcal{K}}\psi_\sigma(q)$ is positive, we still observe cluster formation.

\begin{figure}
	\centering
	\includegraphics[width=0.49\linewidth]{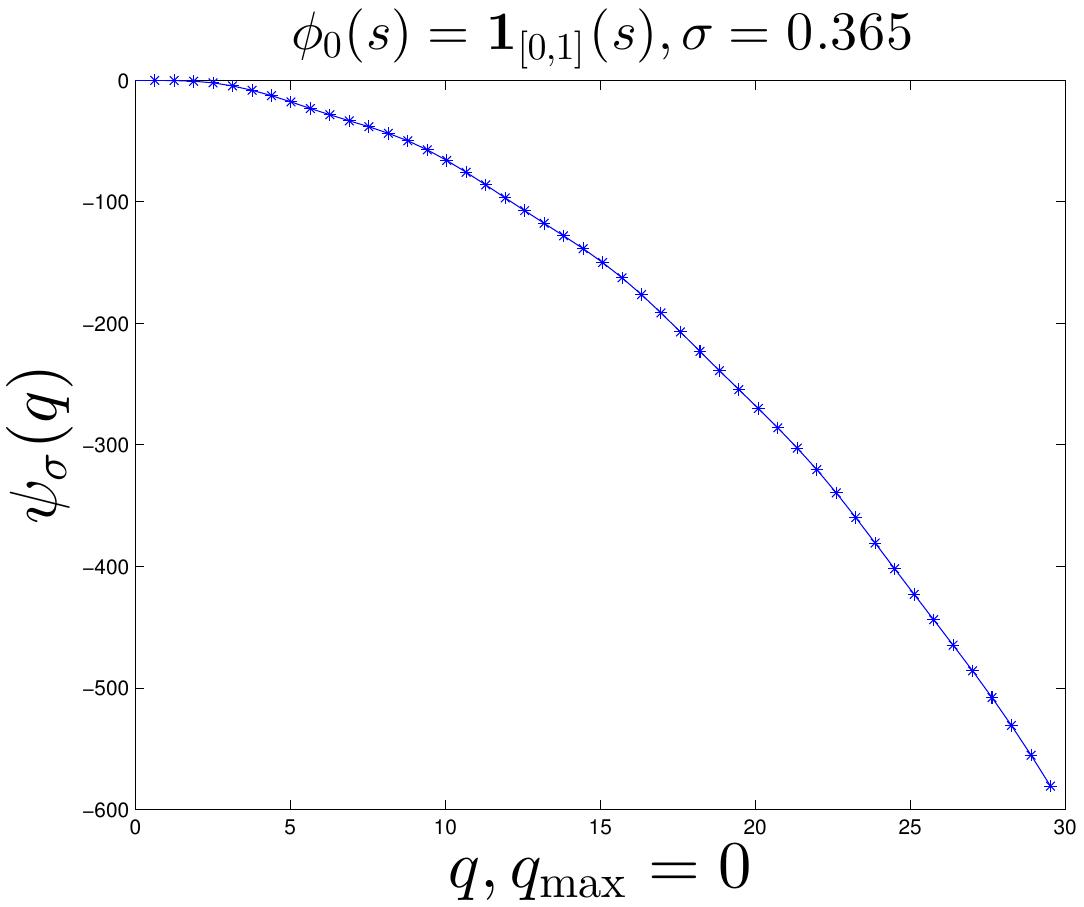}
	\includegraphics[width=0.49\linewidth]{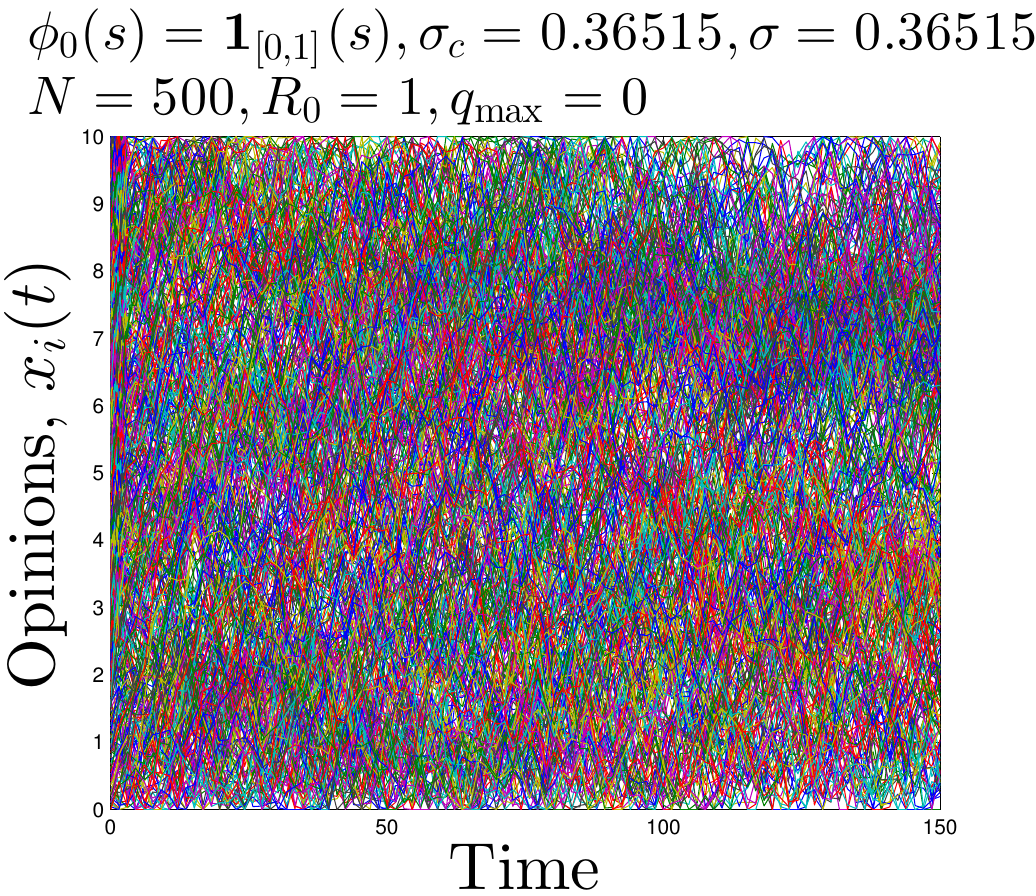}
	\caption{Simulations for $\sigma=\sigma_c$. Left: $\psi_\sigma(q)$ evaluated at $R_0q\in\mathcal{K}$. 
	Right: Simulations of (\ref{eq:stochastic discrete dx_i}).}
	\label{fig:sigma_c for sigmaC}
\end{figure}

\begin{figure}
	\centering
	\includegraphics[width=0.49\linewidth]{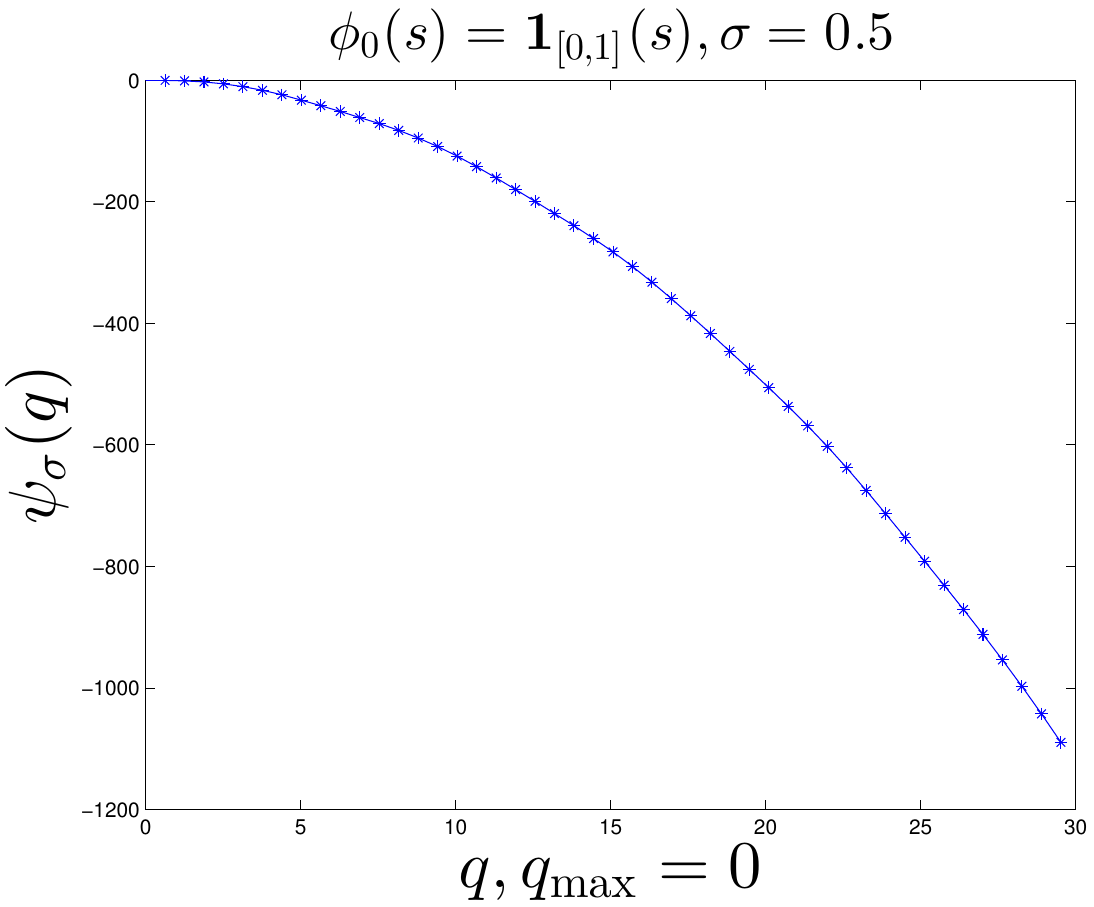}
	\includegraphics[width=0.49\linewidth]{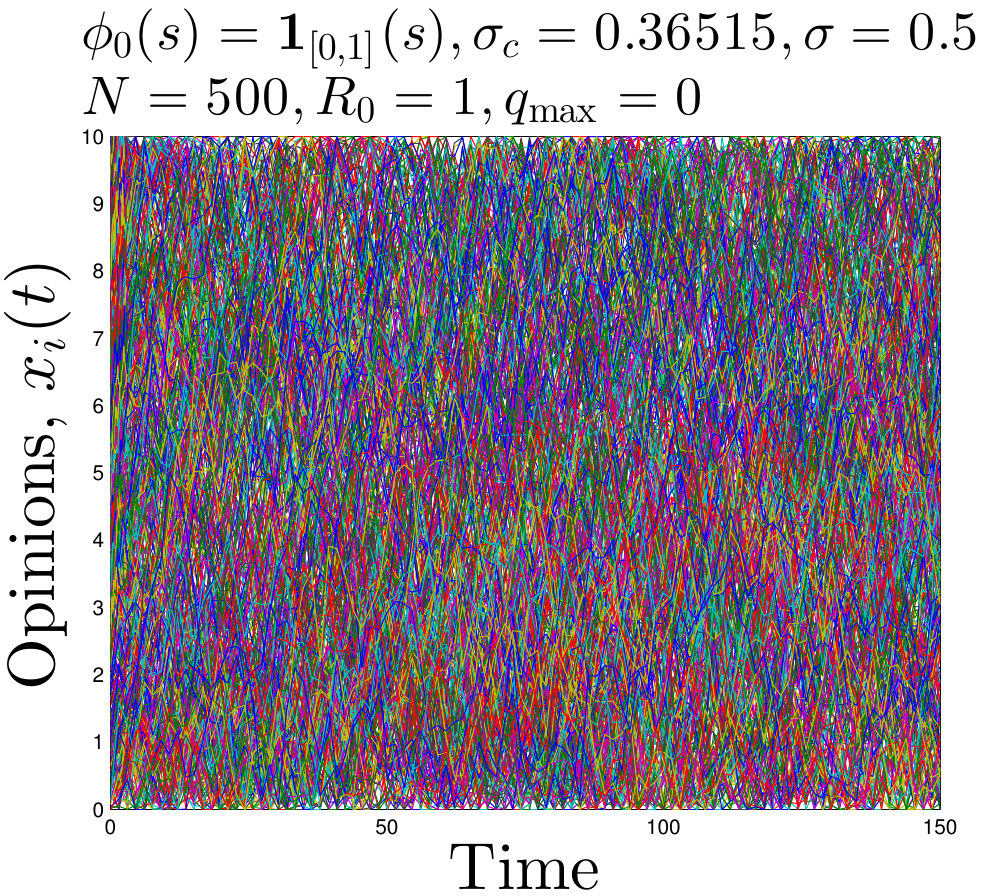}
	\caption{Simulations for $\sigma=0.5$. Left: $\psi_\sigma(q)$ evaluated at $R_0q\in\mathcal{K}$. 
	Right: Simulations of (\ref{eq:stochastic discrete dx_i}).}
	\label{fig:sigma_c for sigma0.5}
\end{figure}

We note that in Figure \ref{fig:sigma_c for sigmaC} and \ref{fig:sigma_c for sigma0.5},  if $\sigma\geq\sigma_c$, $\psi_\sigma(q)<0$ for all 
$q>0$ and $\psi_\sigma(0)=0$. In other words, the linearized system is stable and thus the full system is stable. In this case, we do not see 
cluster formation and the system behaves like an $N$-independent agent system.

We see from the simulations that to model opinion dynamic with consensus convergence it is appropriate to assume that $\sigma<\sigma_c$. 
Therefore we will assume that $\sigma=0.1$ in our simulations of the stochastic system.

We revisit Figure \ref{fig:sigma_c for sigma0.1} to check our analysis. First of all, $q_{\max}=2.5133$ is clearly a unique maximizer and the 
corresponding inter-cluster distance is $2.51$, which agrees with the numerical inter-cluster distance we see in Figure \ref{fig:sigma_c for sigma0.1}. In 
addition, $L/2.51=3.9841$ also predicts well the actual number of the clusters, $4$. Finally, the blue dashed line $t=t_{clu}$ indicates the 
time to the cluster formation, even though Figure \ref{fig:sigma_c for sigma0.1} is just one realization.

We test $t_{clu}$ and the width of clusters in a more statistical way by examining $1000$ realizations. If $\phi_0(s)=\mathbf{1}_{[0,1]}(s)$ and 
$\sigma=0.1$, then we can expect that we will have $4$ clusters at $T=150$ in most of the realizations. For each realization, we numerically 
compute the widths of the clusters, $\hat{\sigma}^{(j)}(t)$, $j=1, \ldots, 4$ by using the empirical standard deviations of 
$\{x^{(j)}_i(t)\}_{j=1}^4$ (see (\ref{eq:density of each cluster})):
\begin{equation}
	\label{eq:empirical standard deviation}
	(\hat{\sigma}^{(j)}(t))^2 = \frac{2}{N^{(j)}-1}\sum_{i=1}^{N^{(j)}}\left(x^{(j)}_i(t)-\bar{x}^{(j)}(t)\right)^2,\quad 
	\bar{x}^{(j)}(t) = \frac{1}{N^{(j)}}\sum_{i=1}^{N^{(j)}} x^{(j)}_i(t)
\end{equation}
where for each $j=1,\ldots,4$, $\{x^{(j)}_i(t)\}$ belong to the $j$-th cluster and $N^{(j)}$ is  the number of agents in the j-th clusters. 
Of course, $\hat{\sigma}^{(j)}(t)$ in (\ref{eq:empirical standard deviation}) is just one realization and so we compute $\hat{\sigma}^{(j)}(t)$ for $1000$ 
realizations and consider the average.

\begin{figure}
	\centering
	\includegraphics[width=0.49\linewidth]{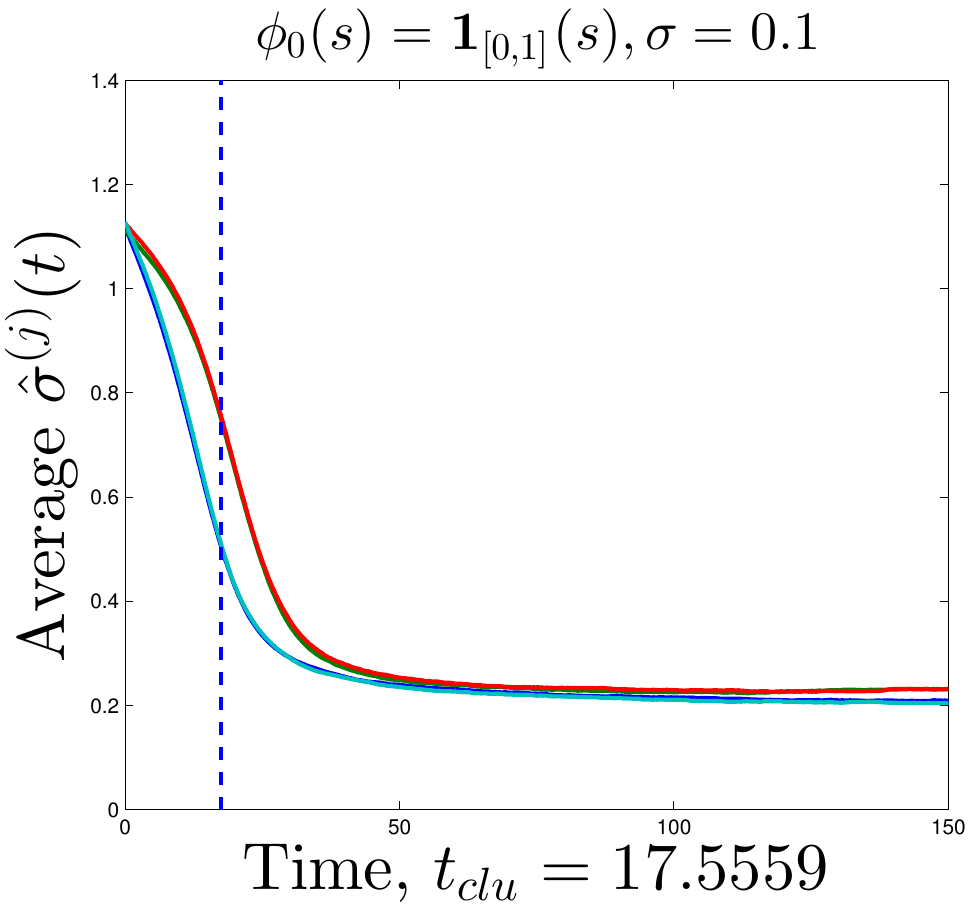}
	\includegraphics[width=0.49\linewidth]{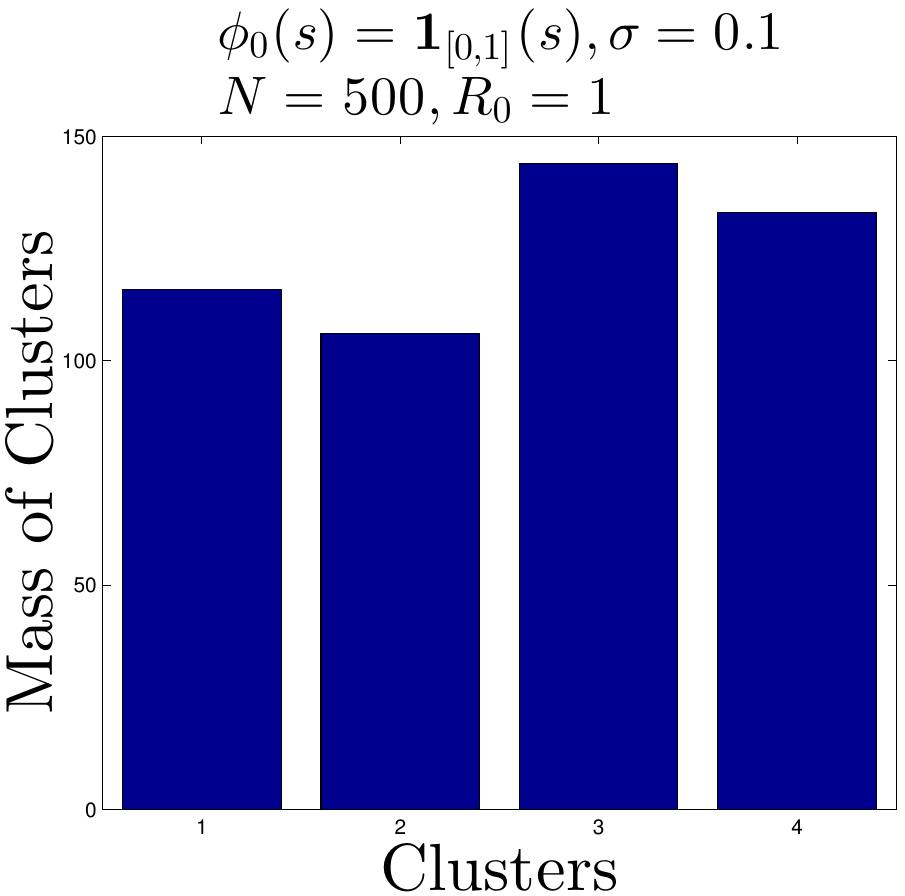}
	\caption{Left: Average of $1000$ realizations of $\hat{\sigma}^{(j)}(t)$, $j=1,\ldots,4$ for $\sigma=0.1$. 
	The vertical dashed line is at $t=t_{clu}$. 
	Right: $N^{(j)}$, the number of agents in the $j$-th clusters in Figure \ref{fig:sigma_c for sigma0.1}.}
	\label{fig:width and mass of clusters}
\end{figure}

The averages $\hat{\sigma}^{(j)}(t)$ are shown in the left part of Figure \ref{fig:width and mass of clusters}, in different colors. First, we can see that $t_{clu}$, as 
expected, is the halfway from the time to maximum with to the time to the minimum width. Second, from (\ref{eq:density of each cluster}), the width of 
each cluster is analytically $\sigma^{(j)} = \sigma/\sqrt{ m_c \phi_0(0)} = 0.1/\sqrt{0.25\times 1}=0.2$, which agrees with the numerical values 
$\hat{\sigma}^{(j)}(t)$ when $t$ is large.

\begin{figure}
	\centering
	\includegraphics[width=0.49\linewidth]{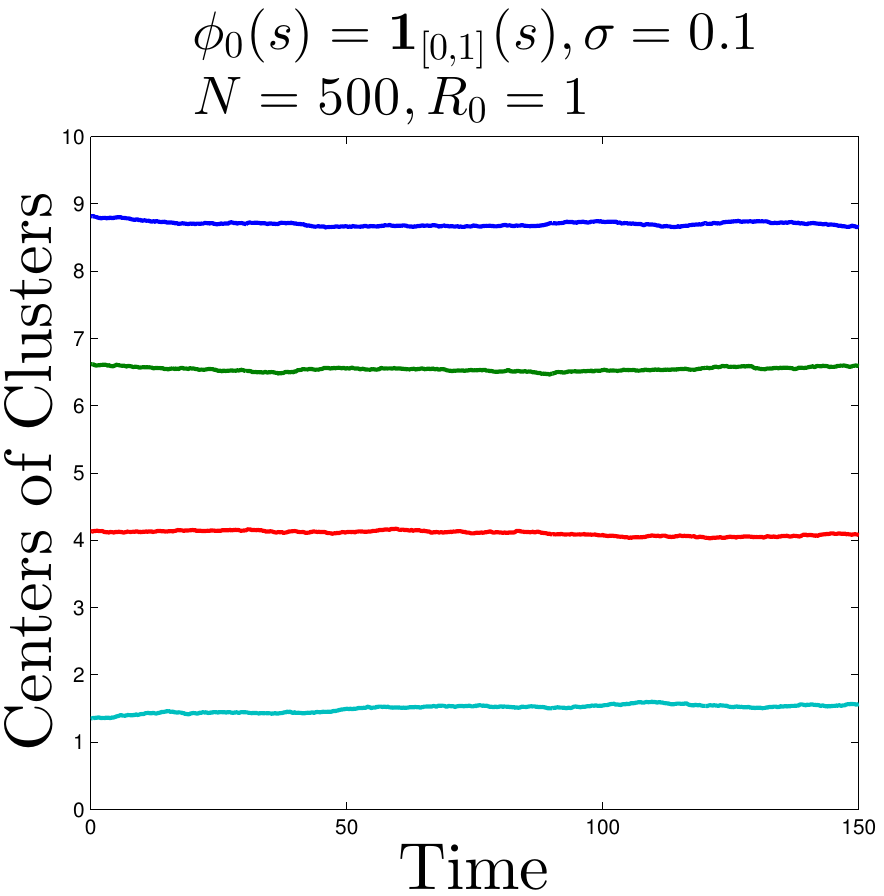}
	\includegraphics[width=0.49\linewidth]{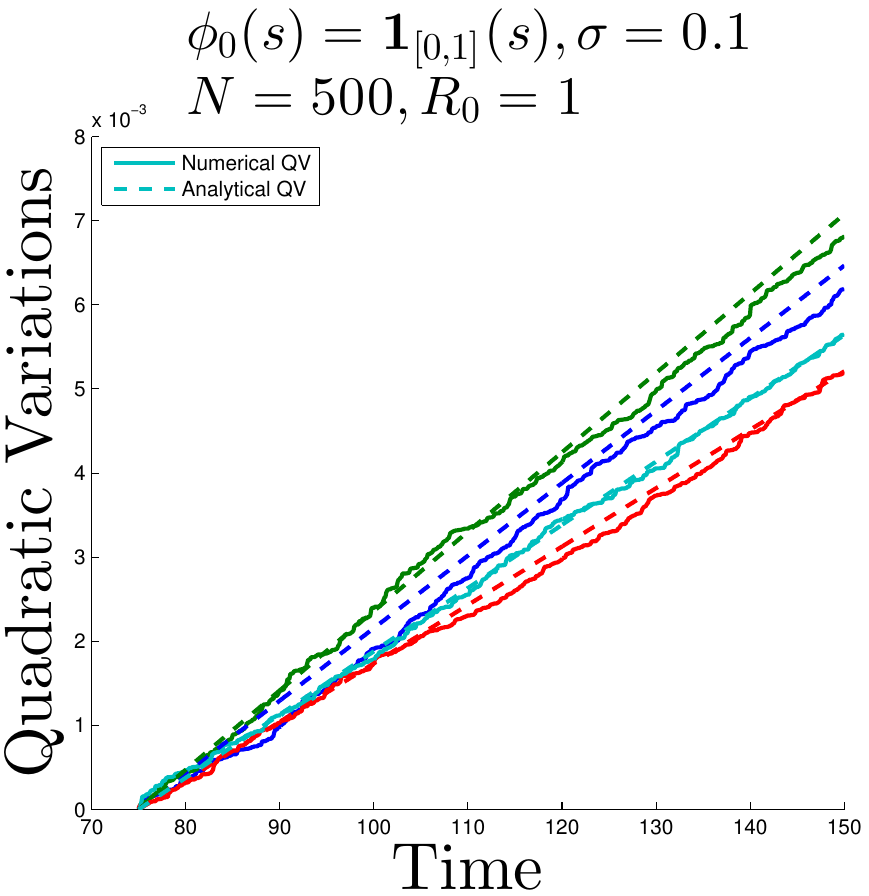}
	\caption{Left: Centers of the clusters, $X^{(j)}(t)$, in Figure \ref{fig:sigma_c for sigma0.1}, in different colors. Right: Quadratic variations of the cluster centers 
	from $t=75$ to $t=150$. Solid lines: Numerical quadratic variations of $X^{(j)}(t)$. Dashed lines: Quadratic variations of 
	$\sigma W^{(j)}(t)/\sqrt{N^{(j)}}$, where $\{W^{(j)}(t)\}_{j=1}^4$ are independent standard Brownian motions and $\{N^{(j)}\}_{j=1}^4$ 
	are the numbers of agents in the clusters.}
	\label{fig:centers and QV}
\end{figure}

We also analyze the behavior of the centers of the clusters. The centers $\{X^{(j)}(t)\}_{j=1}^4$ of the clusters in 
Figure \ref{fig:sigma_c for sigma0.1} are plotted in Figure \ref{fig:centers and QV}. From the previous analysis (\ref{eq:center of cluster}), the 
centers of the clusters are independent Brownian motions $\sigma W^{(j)}(t)/\sqrt{Nm_c}$. For one realization, the opinions $\{x_i(t)\}_{i=1}^N$ 
will not be evenly distributed in the clusters. For example, the actual numbers $\{N^{(j)}\}_{j=1}^N$ of agents of the clusters in 
Figure \ref{fig:sigma_c for sigma0.1} are plotted in Figure \ref{fig:width and mass of clusters}. So for one realization, $X^{(j)}(t)$ is a 
Brownian motion $\sigma W^{(j)}(t)/\sqrt{N^{(j)}}$. On the right part of Figure \ref{fig:centers and QV}, we compare the quadratic 
variations of $X^{(j)}(t)$ and $\sigma W^{(j)}(t)/\sqrt{N^{(j)}}$ for $75\leq t\leq 150$ (after the time to the cluster formation.) Indeed, 
from the figure, we can see that their quadratic variations are very similar and that means $X^{(j)}(t)$ are very close to Brownian motions.

\section{Long time behavior of simulations}
\label{sec:long}

\begin{figure}
	\centering
	\includegraphics[width=0.49\linewidth]{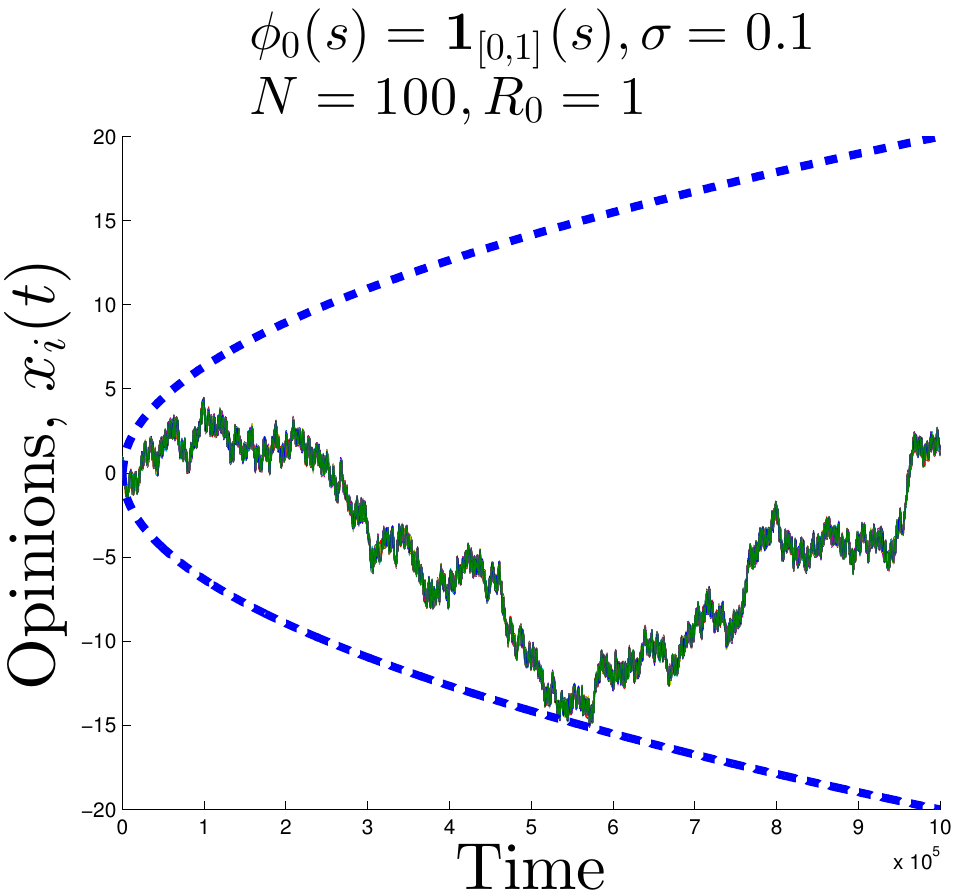}
	\includegraphics[width=0.49\linewidth]{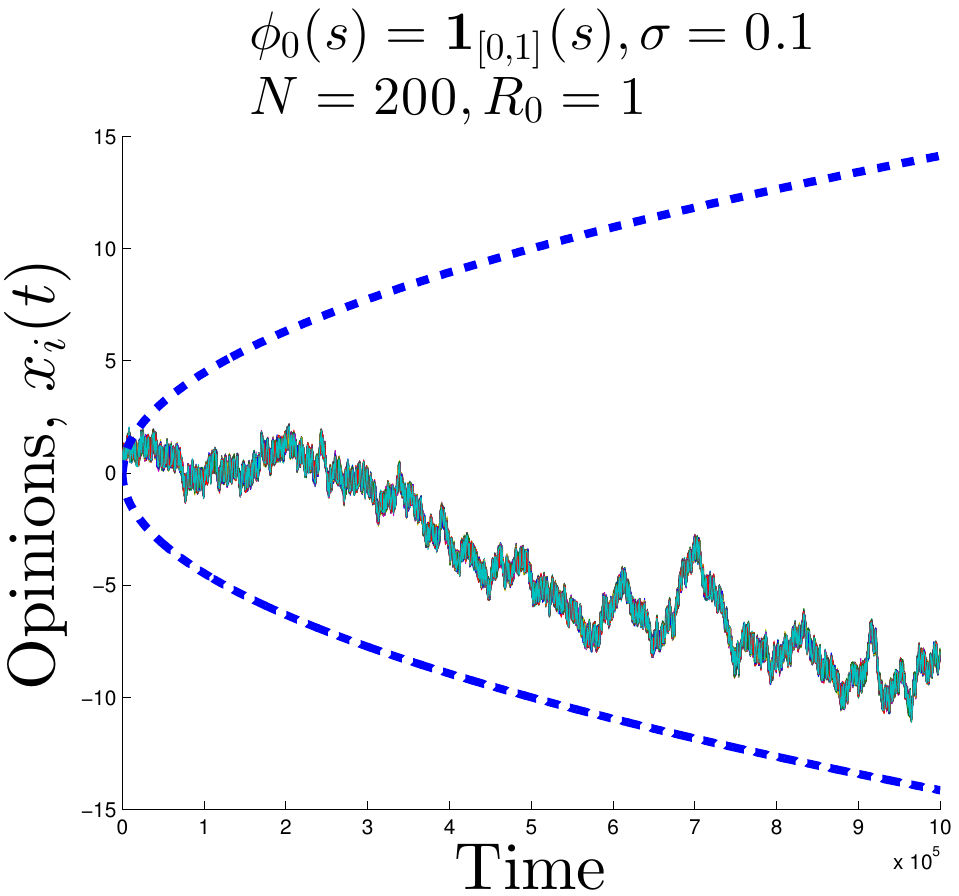}
	\caption{Long time behavior of the opinions for $\sigma=0.1$ and for $N=100, 200$. The blue dashed curve is the equation 
	$x=\pm2\sigma\sqrt{t/N}$. $\sigma$ is small and the overall behavior is like a single Brownian motion.}
	\label{fig:long time, sigma 0.1}
\end{figure}

\begin{figure}
	\centering
	\includegraphics[width=0.49\linewidth]{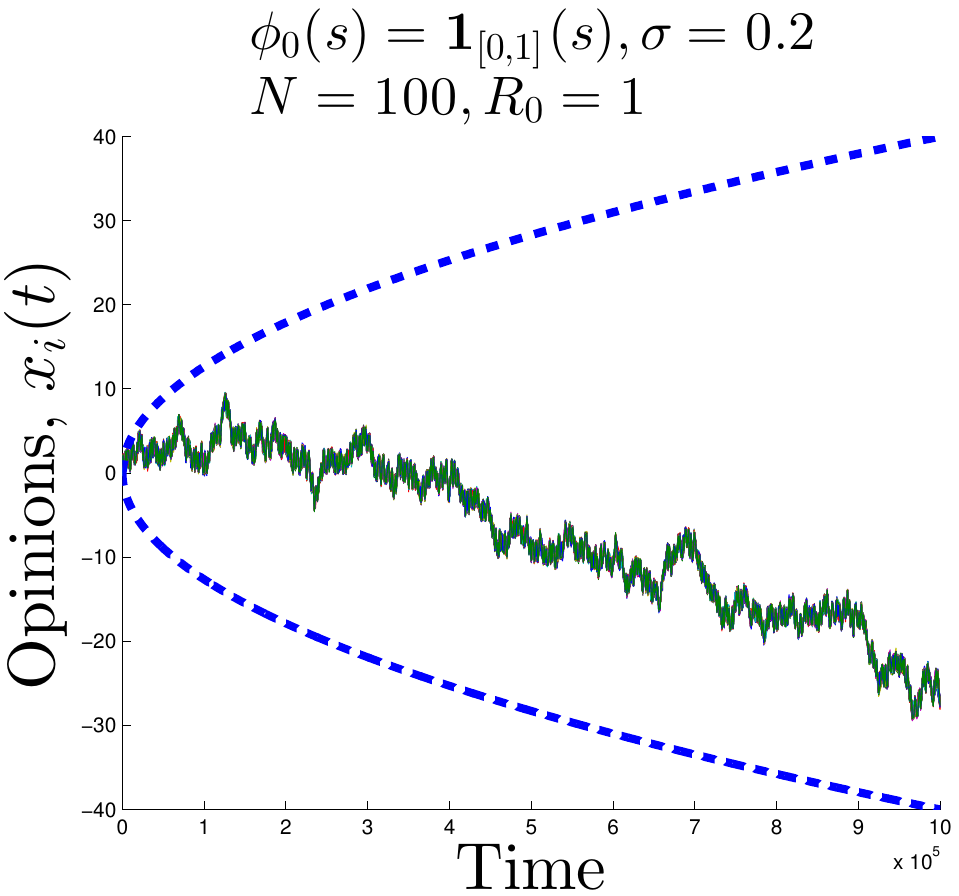}
	\includegraphics[width=0.49\linewidth]{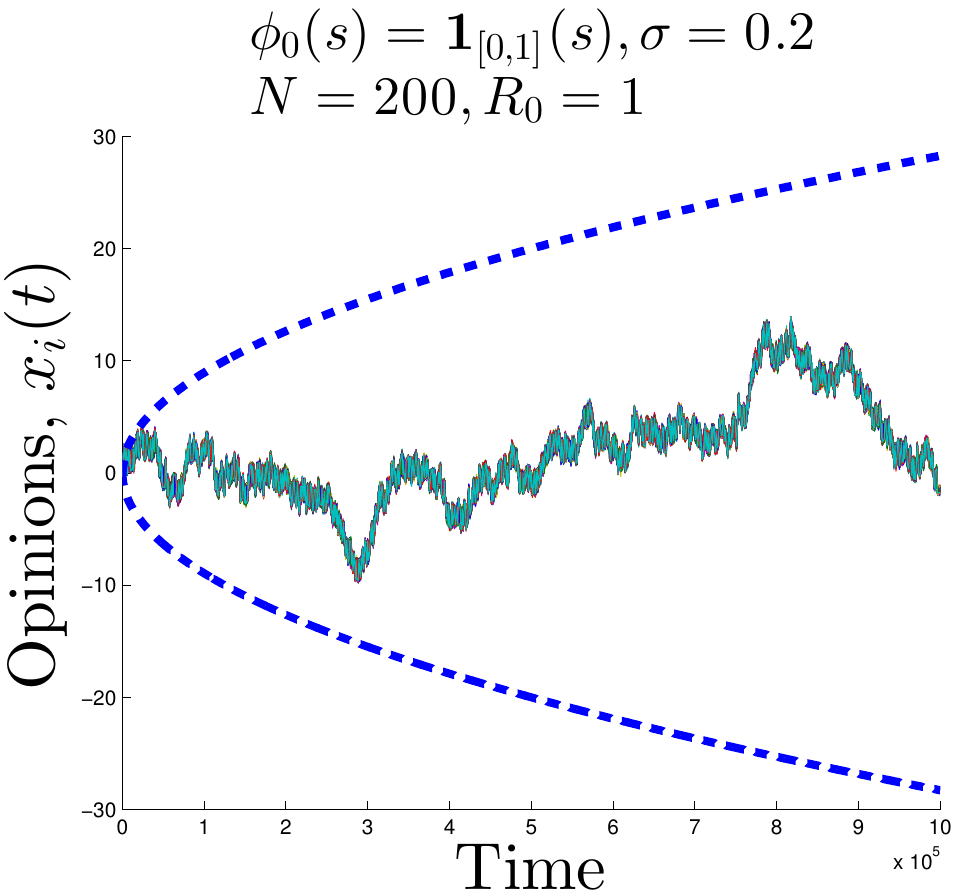}
	\caption{Long time behavior of the opinions for $\sigma=0.2$ and for $N=100, 200$. The blue dashed curve is the equation 
	$x=\pm2\sigma\sqrt{t/N}$. $\sigma$ is small and the overall behavior is like a single Brownian motion.}
	\label{fig:long time, sigma 0.2}
\end{figure}

\begin{figure}
	\centering
	\includegraphics[width=0.49\linewidth]{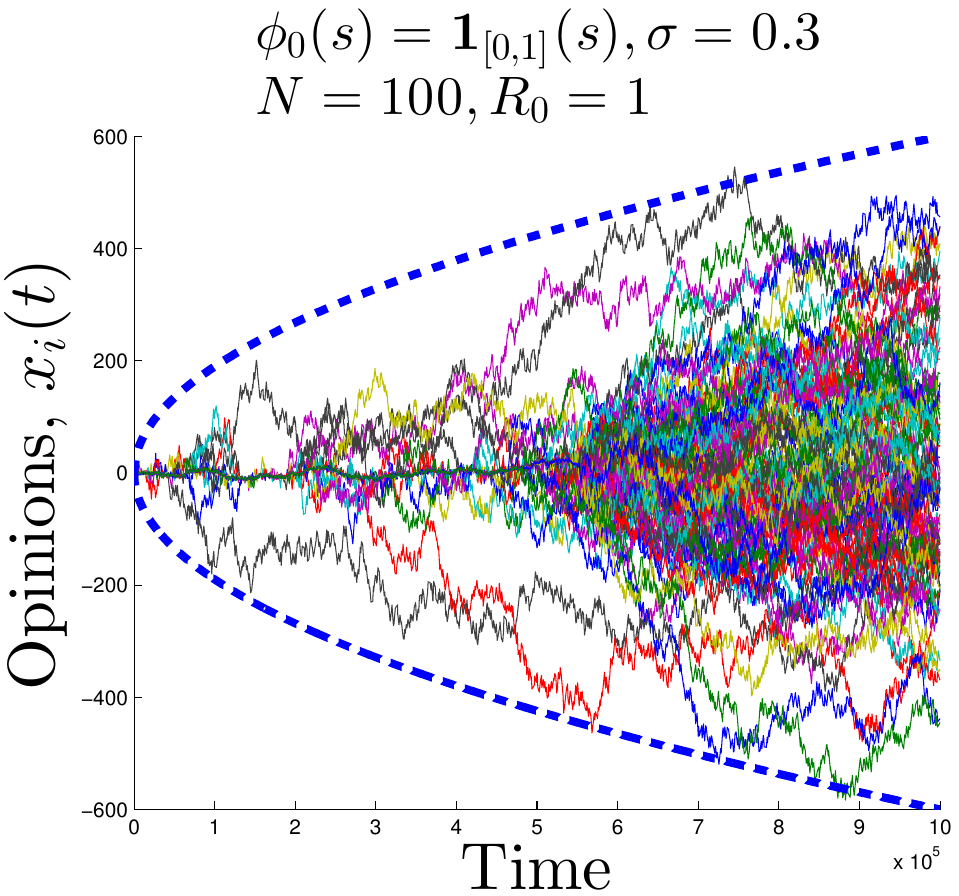}
	\includegraphics[width=0.49\linewidth]{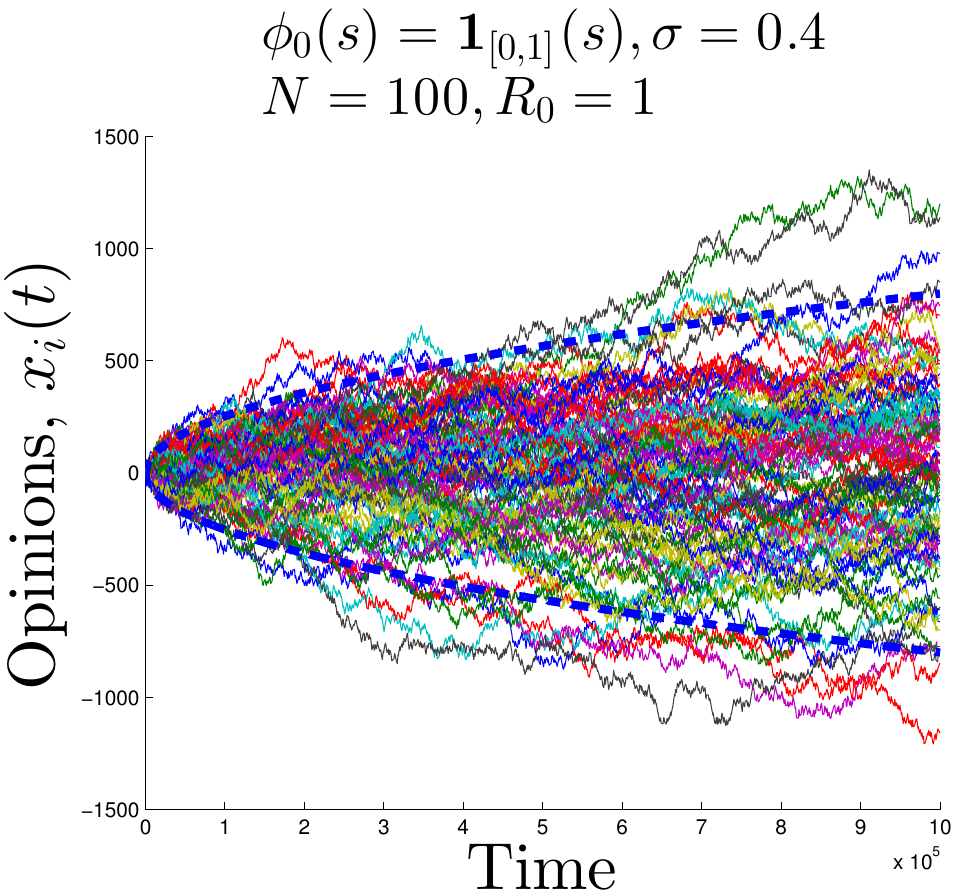}
	\caption{Long time behavior of the opinions for $\sigma=0.3, 0.4$ and for $N=100$. The blue dashed curve is the equation 
	$x=\pm2\sigma\sqrt{t}$. For $\sigma=0.3$, there is a single cluster for $t\leq 5.5\times10^5$. However, for $t>5.5\times10^5$, $x_i(t)$
	disintegrate and the system behaves like an $N$-independent Brownian motions. For $\sigma=0.4$, the random perturbations are large enough so 
	that the system is an $N$-independent Brownian motions at the beginning.}
	\label{fig:long time, sigma 0.3 and 0.4}
\end{figure}

We have also simulated numerically the long time behavior of the system defined on the full real line $\mathbb{R}$, 
especially the behavior after the onset of consensus convergence.  As we discuss in 
the previous section, when there is randomness the center of the unique cluster behaves like a diffusion process $\sigma W(t)/\sqrt{N}$, where $W(t)$ is a standard 
Brownian motion. In Figure \ref{fig:long time, sigma 0.1} and \ref{fig:long time, sigma 0.2}, we observe that the centers indeed behave like Brownian motions.  The 
dashed lines are the parabolas with equation $x=\pm2\sigma\sqrt{t/N}$ so that for any fixed $t$, the centers are within the parabolas with $95\%$ probability.

However, when $\sigma$ is sufficiently large, the long time behavior is different. On the right part of Figure \ref{fig:long time, sigma 0.3 and 0.4}, when 
$\sigma=0.4>\sigma_c$, the system behaves like $N$-independent diffusions.  A more interesting case is when $\sigma=0.3<\sigma_c$ on the left part of Figure 
\ref{fig:long time, sigma 0.3 and 0.4}.  In this case, for $0\leq t\leq 5.5\times10^5$ there is still consensus convergence, but for $t>5.5\times10^5$, all 
$x_i(t)$ spread out from the unique cluster and the system becomes an independent agent evolution.  A detailed mathematical analysis using large deviations theory
for such a phenomenon is  being considered at present.
\section{Conclusion}
\label{sec:conclusion}

We have analyzed a stochastic, continuous time opinion dynamics model and we have carried out extensive numerical simulations. 
We use the mean-field theory and obtain a nonlinear 
Fokker-Planck equation as the number of opinions tends to infinity.  Then we use a linear stability analysis to estimate the critical value of the noise strength  
so as to have cluster formation, estimate the number of clusters and the time to cluster formation. These quantities are closely related to the frequency that maximizes the 
growth rate of the linearized modes (\ref{eq:stochastic gamma_k}).  After the initial cluster formation we expect, and numerically confirm, that the centers of the clusters 
behave like Brownian motions before further consolidation.  
Finally, the long time behavior of the system is explored numerically and we observe that after a unique cluster is formed, there 
is a small probability that the opinions will spread out from the unique cluster and the system will become an independent agent evolution.

\bibliographystyle{plain}
\bibliography{reference}
\end{document}